\documentclass[a4paper, notitlepage, 10pt]{article}
\usepackage{geometry}
\usepackage{titling}
\usepackage{authblk}

\usepackage{setspace,relsize}
\fontfamily{times}
\usepackage{moreverb}

\usepackage{booktabs}
\usepackage{graphicx}
\usepackage{subcaption}
\usepackage{tikz}
\usetikzlibrary{shapes}
\usepackage{wrapfig}
\usepackage{listings}

\usepackage{amsthm}
\usepackage{amsmath}
\usepackage{amssymb}
\usepackage{mathtools}
\usepackage{bbm}

\usepackage{indentfirst}
\usepackage{multirow}
\usepackage{enumitem}
\usepackage{cancel}
\usepackage[normalem]{ulem}

\definecolor{keywords}{RGB}{255,0,90}
\definecolor{comments}{RGB}{0,0,113}
\definecolor{red}{RGB}{160,0,0}
\definecolor{green}{RGB}{0,150,0}

\usepackage{hyperref}
\usepackage{xcolor}
\hypersetup{
    colorlinks=true,
    linkcolor=blue,      
    citecolor=blue,      
    urlcolor=blue        
}
\usepackage[square,numbers]{natbib}

\title{\Large \bfseries Optimal vaccination strategies on networks and in metropolitan areas}
\date{\today}
\author[1]{Lucas Machado Moschen}
\author[2]{María Soledad Aronna}
\affil[1]{\small Sorbonne University}
\affil[2]{\small School of Applied Mathematics, Fundação Getulio Vargas}

\newcommand{\R}{\mathbb{R}}
\newcommand{\bb}[1]{\boldsymbol{#1}}

\newcommand{\almost}{\text{a.e. }}

\newcommand{\rzero}{\mathcal{R}_0}
\newcommand{\Di}{v^{\text{max}}_i}

\newtheorem{theorem}{Theorem}[section]

\newtheorem{proposition}{Proposition}[section]
\newtheorem{corollary}{Corollary}[section]
\newtheorem{lemma}{Lemma}[section]

\theoremstyle{definition}

\theoremstyle{definition}

\theoremstyle{remark}
\newtheorem{remark}{Remark}[section]
\newtheorem{assumption}{Assumption}[section]

\begin{document}
\maketitle

\begin{abstract}
This study presents a mathematical model for optimal vaccination strategies in interconnected metropolitan areas, considering commuting patterns. 
It is a compartmental model with a vaccination rate for each city, acting as a control function. 
The commuting patterns are incorporated through a weighted adjacency matrix and a parameter that selects day and night periods. 
The optimal control problem is formulated to minimize a functional cost that balances the number of hospitalizations and vaccines, including restrictions of a weekly availability cap and an application capacity of vaccines per unit of time.
The key findings of this work are bounds for the basic reproduction number, particularly in the case of a metropolitan area, and the study of the optimal control problem.
Theoretical analysis and numerical simulations provide insights into disease dynamics and the effectiveness of control measures.
The research highlights the importance of prioritizing vaccination in the capital to better control the disease spread, as we depicted in our numerical simulations.
This model serves as a tool to improve resource allocation in epidemic control across metropolitan regions. 

\end{abstract}

\section{Introduction}

Metropolitan areas consist of a densely populated urban core, such as a capital city, along with its surrounding territories that share social, economic, and infrastructural ties. 
One defining characteristic of these regions is commuting: the routine movement of individuals between their city of residence and the city of the workplace.
For instance, in the Rio de Janeiro Metropolitan Area, over two million people commute daily, in a 13-million inhabitants region~\cite{sebrae2013}. 
Such mobility creates a complex network of interconnected cities, where the links are weighted by the flux of people.
The recent global health outbreaks, such as the Zika virus epidemic and the COVID-19 pandemic, have shown how these dense and interconnected urban areas can amplify disease transmissibility.
Therefore, devising strategies that efficiently reduce the impact of outbreaks is essential, and mathematical modeling serves as an indispensable tool to perform this rigorously.

Mathematical Epidemiology has been instrumental in understanding how social interactions and human mobility influence outbreaks.
The main tool in this field is compartmental models, which divide the population into groups based on their disease status ({\em e.g.,} susceptible, infectious, recovered) with differential equations representing the dynamics.
They can integrate various real-life factors, including age, spatial effects, etc.
The inclusion of spatial effects is particularly relevant as the transmission of infectious diseases often depends on proximity and movement patterns of individuals.
The spatial aspect can be approached in either continuous or discrete ways.
The latter, known as the {\em patch\/} or {\em metapopulation model}, first divides the population into distinct subpopulations, then further into compartments, resulting in a large set of equations, one for each compartment.
Introduced in ecology to study competitive species~\cite{levin1993patch}, this approach gained prominence in epidemiology after the work of \citeauthor{rvachev1985mathematical} (\citeyear{rvachev1985mathematical}), who used the airplane network to predict the spread of influenza, focusing on the 1968 Hong Kong epidemic~\cite{rvachev1985mathematical}.

Building on this foundational work, \citeauthor{sattenspiel1995structured} (\citeyear{sattenspiel1995structured}) and \citeauthor{arino2003multi} (\citeyear{arino2003multi}) discussed the integration of mobility between regions into epidemic models, with the latter arguing in favor of space-discrete models and obtaining inequalities for the basic reproduction number $\rzero$~\cite{arino2003multi, sattenspiel1995structured}. 
Extending the approaches into epidemic propagation, \citeauthor{takeuchi2006spreading} (\citeyear{takeuchi2006spreading}) analyzed the impact of transport-related infections on the disease transmission, being followed by Liu et al.~\cite{liu2009global, liu2013transmission}. 
Considering the structure of the graph associated with the city's connectivity, \citeauthor{colizza2008epidemic}~\cite{colizza2008epidemic, colizza2007reaction} calculated a global threshold for disease invasion and provided extensive Monte Carlo simulations to verify their findings.
They also analyzed the effect of the topology of the graph on the phase diagram in metapopulation models.
The works of~\citeauthor{pastor2015epidemic} \cite{pastor2001epidemic, pastor2002epidemic, pastor2015epidemic} were also prominent in understanding infectious diseases in networks of individuals.
We refer the reader to~\cite{yin_novel_2020} to see the combination of metapopulation and agent-based models for two cities.
\citeauthor{stolerman2015sir} (\citeyear{stolerman2015sir}) introduced an SIR-network model to understand the impact of infection rates on $\rzero$, inferring that nodes with higher infection rates are the most important drivers of outbreaks~\cite{stolerman2015sir}. 

To effectively tame the epidemic's outbreak,  optimal control theory has shown to be useful for theoretical basis and practical solutions~\cite{brauer2012mathematical, lenhart2007optimal, sharomi2017optimal}.
In optimal control formulations, one usually aims at minimizing the number of infections, deaths, or other epidemic-related quantity, at the expense of some control measure, such as quarantine, testing, treatment, and/or vaccination.
Several works came in this sense, such as \citeauthor{behncke2000optimal} (\citeyear{behncke2000optimal})~\cite{behncke2000optimal} who proved that the optimal strategy is to vaccinate as many people as possible as quickly as possible, subject to the resources.
See also \cite{gaff2009optimal, neilan2010introduction} for other related references. 
The interplay of optimal control theory with metapopulation models in epidemiology has contributions from \citeauthor{ogren2002vaccination} (\citeyear{ogren2002vaccination}), \citeauthor{asano2008optimal} (\citeyear{asano2008optimal}), \citeauthor{rowthorn2009optimal} (\citeyear{rowthorn2009optimal}), with increasing traction after the COVID-19 pandemic.
The focus of these studies was to improve vaccination strategies, especially in urbanized, mobile populations, and provide solutions tailored to economic constraints and quarantined settings~\cite{asano2008optimal, ogren2002vaccination, rowthorn2009optimal}.

However, most optimal control applications in epidemiology involved only simple constraints, while the interest in including real-world restrictions has grown.
\citeauthor{hansen2011optimal} (\citeyear{hansen2011optimal}) notably included resource constraints in their SIR models. 
These advancements in constrained optimal control problems were studied by \citeauthor{biswas2014seir} (\citeyear{biswas2014seir}) and \citeauthor{de2015optimal} (\citeyear{de2015optimal}). 
The COVID-19 pandemic accelerated these developments, with research like that of \citeauthor{avram2022optimal} (\citeyear{avram2022optimal}) integrating constraints to simulate the capacity of intensive care units.
\citeauthor{lemaitre2022optimal} (\citeyear{lemaitre2022optimal})~\cite{lemaitre2022optimal} developed a method that combines distributed direct multiple shooting, automatic differentiation, and large-scale nonlinear programming to optimally allocate COVID-19 vaccines, considering both supply and logistic constraints. 
When applying their approach to cities in Italy, they used a SEPIAHQRD-V model. 
They specified that individuals in certain compartments do not commute and accounted for mobility fluxes and infection forces that depend on each region.
Another work that emerged to deal with coronavirus was the Robot Dance platform, developed by \citeauthor{nonato_robot_2022} (\citeyear{nonato_robot_2022}). 
This platform is a computational tool aiding policymakers in curating response strategies tailored to regional nuances, intercity commuting mobility, and hospital capacities, applied to the state of São Paulo, Brazil~\cite{nonato_robot_2022}.

In this work, we delve into the intricate dynamics of epidemic spread in metropolitan areas, emphasizing the role of commuting patterns and vaccination strategies. 
We derive tight bounds for the basic reproduction number,  for both a general network of cities and one associated with a metropolitan area.
Utilizing optimal control theory, we devise efficient strategies for disease control in metropolitan areas, which we illustrate through a series of numerical experiments that corroborate our findings.

\subsection{Contributions}\label{sec:contributions}

In this work, we study a mathematical model that combines commuting and vaccination in a constrained optimal control problem to manage the propagation of infectious diseases within metropolitan areas, providing a more realistic representation of disease transmission and an efficient strategy to control it.
Our main contributions are three-fold:

\begin{itemize}
    \item Analysis of a mathematical model for epidemics that considers commuting dynamics in metropolitan areas and derivation of upper and lower bounds for the basic reproduction number.
    
    \item Formulation of a constrained control-affine optimization problem for finding optimal vaccination strategies in a network of cities.
    
    \item Extensive numerical simulations, illustrating various situations and validating the theoretical results.
\end{itemize}

\section{Methods}
\label{sec:methods}

The formulation of the vaccination model and the associated optimal control problem is composed of three elements: (I) the epidemiological model that combines a compartmental SIR model and commuting patterns to describe the transmission dynamics; 
(II) an objective function to be minimized, which contemplates  the number of infections and applied vaccines; 
(III) a set of constraints on the state and control variables that ensure resource limitations for the production, distribution, and application of the vaccines.

\subsection{Epidemiological modeling}

Consider a directed graph $ G = (V, E) $, where the set of nodes $V$ corresponds to $K$ interconnected cities, and the set of ordered pairs $E$ indicates connectivity between cities. 
Each link $(i,j)$ has a weight $p_{ij}$ denoting the fraction of the population of city $i$ that commutes daily to the city $j$. 
This forms a {\em weighted adjacency matrix}  ${[P]}_{ij} = p_{ij}$, with $P$ being a right stochastic matrix, {\em e.g.} the elements of each row sum up $1$. 
Each city $i$ has a total population $n_i$ divided into three compartments representing the proportion of susceptible ($S_i$), infectious ($I_i$), and recovered ($R_i$) individuals, satisfying:
\[
S_i(t) + I_i(t) + R_i(t) = 1,
\]
for all $ t \ge 0 $ and all $i=1,\dots,K$.

In this model, we incorporate a birth rate $\mu$ and a natural death rate $\mu$ for each compartment in each city. 
The rates are chosen such that the total population of each city remains constant over time.
We also introduce $\alpha$, a daily periodic function, such that $\alpha(t) = 0$ during daytime and $\alpha(t) = 1$ at night. For instance, a realistic setting for $\alpha$ could be $\alpha(t) = 0$ for $t\in (k+1/4,k+3/4),$ and $\alpha(t) =1$ for $t \notin (k+1/4,k+3/4)$, for each day $k=0,1,\dots,T$.
During nighttime, the evolution of the system is not influenced by mobility, leading to the SIR model:
\[
\begin{aligned}
    \frac{dS_i}{dt}(t) &= \mu - \beta_i S_i I_i - \mu S_i, \\
    \frac{dI_i}{dt}(t) &= \beta_i S_i I_i - \gamma I_i - \mu I_i, \\
    \frac{dR_i}{dt}(t) &= \gamma I_i - \mu R_i,
\end{aligned}
\]
where $\beta_i$ represents city $i$'s infection rate, which is directly affected by the population density and the probability of infection given contact, and $\gamma^{-1}$ refers to the average infection period.

During the day, commuting changes the infection dynamics. We model this feature using the same approach as \citeauthor{nonato_robot_2022} (\citeyear{nonato_robot_2022})~\cite{nonato_robot_2022}. 
The {\em effective} population in city $i$ during working hours is given by
\[
P_i^{\mathrm{eff}} = \sum_{j=1}^K p_{ji} n_j,
\]
which sums over the parcels of the populations that travel from any city $j$ to $i$. 
The proportion of susceptible workers from city $i$ that get exposed to the infectious individuals in city $j$ is $p_{ij} S_i I_j^{\mathrm{eff}}$, where $I_j^{\mathrm{eff}}$ is the effective proportion of infectious people in city $j$ during the day and it is defined by
\[
I_j^{\mathrm{eff}} = \frac{1}{P_j^{\mathrm{eff}}} \sum_{k=1}^K p_{kj} I_k n_k.
\]

\begin{figure}[!thbp]
    \centering
    \includegraphics[width=\textwidth]{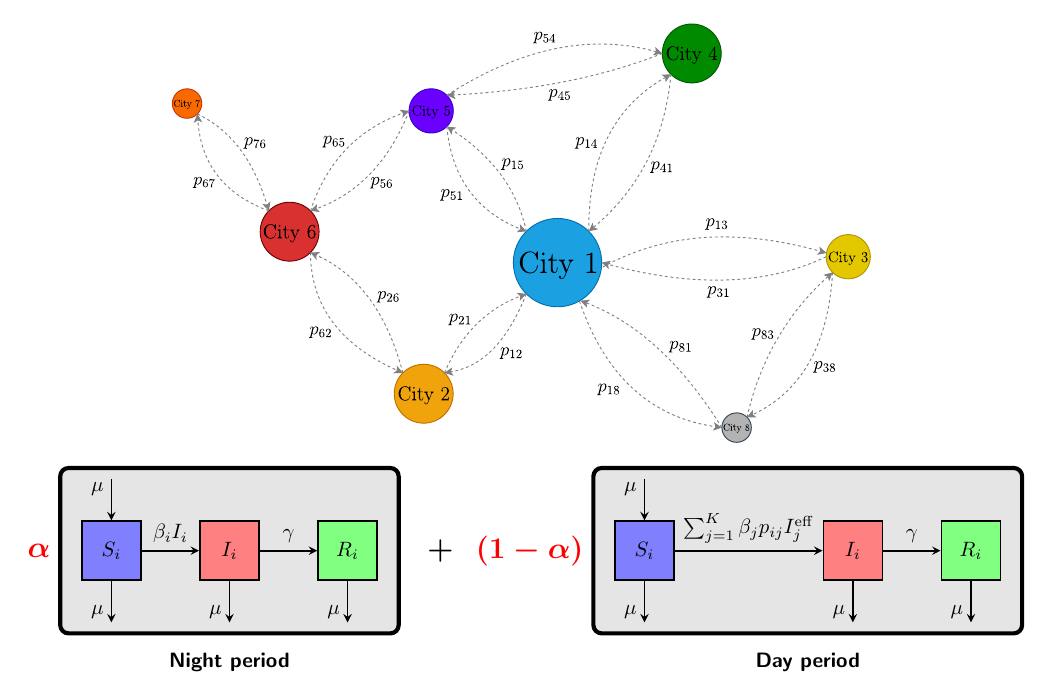}
    \caption{\textbf{Graphical representation of city interactions and the SIR model.}\\
    The cities are connected composing a network, $p_{ij}$ being the proportion of individuals from city $i$ who work during the day at $j$.
    At night, cities do not interact, and the epidemic follows a standard SIR dynamic.
    During the day, some individuals commute, interact with infected individuals in the destination city, and may transport the infection.
    }\label{fig:cities}
\end{figure}

The complete model is
\begin{equation}
    \begin{aligned}\label{eq:robot-dance-model}
    &\frac{dS_i}{dt}(t) = \mu -\alpha(t) \beta_i S_i I_i - (1-\alpha(t)) S_i \sum_{j=1}^K \beta_j p_{ij} I_j^{\mathrm{eff}} - \mu S_i, \\
    &\frac{dI_i}{dt}(t) = \alpha(t) \beta_i S_i I_i + (1-\alpha(t)) S_i \sum_{j=1}^K \beta_j p_{ij} I_j^{\mathrm{eff}} - \gamma I_i - \mu I_i, \\
    &\frac{dR_i}{dt}(t) = \gamma I_i - \mu R_i,
\end{aligned}
\end{equation}
for $i=1,\dots,K$, subject to the initial conditions $S_i(0) = S_{i0}, I_i(0) = I_{i0}$ and $R_i(0) = R_{i0}$, for $i=1,\dots,k$.
We use the following vector notations $\bb{S}(t) = (S_1(t), \dots, S_K(t)) \in \R^K$, $\bb{I}(t) = (I_1(t), \dots, I_K(t)) \in \R^K$ and  $\bb{R}(t) = (R_1(t), \dots, R_K(t)) 
\in \R^K$.
\autoref{fig:cities} represents this model graphically.

\begin{remark}
    In this study, we do not consider disease-specific mortality. 
    This decision is based on two observations: firstly, including an additional mortality rate due to the disease does not change the disease dynamics. 
    Secondly, our primary focus lies in analyzing the impact of vaccination strategies during the initial months of the epidemic, a period where the relative impact of disease-specific mortality on the outcomes is usually minimal. 
    Our main objective is to control the rate of new infections, which is not influenced by the mortality rate.
\end{remark}

It can be proved that 
 \[
    \mathcal{C} \coloneqq \{X \in \R_{\geq 0}^{K \times 3} : X_{i1} + X_{i2} + X_{i3} = 1, \text{ for } i=1,\dots,K\}    
\]
is invariant under system \eqref{eq:robot-dance-model}. The complete statement and its proof are in \autoref{sec:appendix-proof}.

\subsection{The basic reproduction number}\label{sec:basic_reproduction_number}

Following the method developed by Van den Driessche and Watmough \cite{van2002reproduction}, we derive the basic reproduction number using the formula $\rzero = \rho(FV^{-1})$. 
Here $F$ and $V$ are matrices associated with the Jacobian of the dynamics at the disease-free equilibrium (DFE). 
The detailed definitions are included in the \autoref{sec:r0_general_models} and the derivation is given in \autoref{sec:appendix-r0}.
Given the problem's dimension, we could not obtain an explicit expression for $\rzero$. However, we obtained in \autoref{thm:bounds_rzero_generalP}, \autoref{cor:lower_upper_bound} and \autoref{thm:bounds_rzero_metropolitan} below, sharp inequalities for $\mathcal{R}_0$ in terms of simple expressions involving the parameters.

Let $\rzero^i$ be the basic reproduction number of city $i$ if the cities were isolated. 
This is given by the well-known formula
\begin{equation}\label{eq:isolated_cities}
    \rzero^i = \frac{\beta_i}{\gamma + \mu}.
\end{equation}
We get the following result, the proof of which is in \autoref{sec:appendix-proof} .

\begin{theorem}\label{thm:bounds_rzero_generalP}
    The basic reproduction number $\rzero$ for system~\eqref{eq:robot-dance-model} satisfies the following inequalities
    \begin{equation}
        \label{eq:rzero-inequality}
        \min_{1 \le i \le K} \alpha \rzero^i + (1-\alpha) \sum_{k=1}^K p_{ik} \rzero^k \le \rzero \le \max_{1 \le i \le K}  \alpha \rzero^i + (1-\alpha) \sum_{k=1}^K p_{ik} \rzero^k.   
    \end{equation}
\end{theorem}

The inequality in~\eqref{eq:rzero-inequality} can be written as 
\[
\min_{1 \le i \le K} w_i \le \rzero \le \max_{1 \le i \le K} w_i,\quad  \text{ where } w \coloneqq \frac{(\alpha I + (1- \alpha) P)\boldsymbol{\beta}}{\gamma+\mu} \in \mathbb{R}^K,
\]
in which $\boldsymbol{\beta} = (\beta_1, \dots, \beta_K)$.
The matrix $\alpha I + (1-\alpha) P$ is right stochastic and balances the static state, where people are in their residence cities, with probability $\alpha$, and the transition matrix $P,$ that represents commuting, with probability $1-\alpha$.
Therefore, the entry $w_i$ represents the expected basic reproduction number for city $i$, calculated as the average of $\rzero^j$ values for all cities $j$, with the weights determined by the probability of individuals from city $i$ work in city $j$. 

\begin{corollary}\label{cor:lower_upper_bound}
    The following inequality holds: 
    \[
    \min_{1 \le i \le K} \rzero^i \le \rzero \le \max_{1 \le i \le K} \rzero^i.
    \]
    In particular, if we consider, without loss of generality, that $\displaystyle\beta_1 = \max_{1 \le i \le K} \beta_i$, we have that $\rzero \le \rzero^1$.
\end{corollary}

The proof is in \autoref{sec:appendix-proof}.

\begin{figure}[!htbp]
    \centering
    \includegraphics[width=0.8\textwidth]{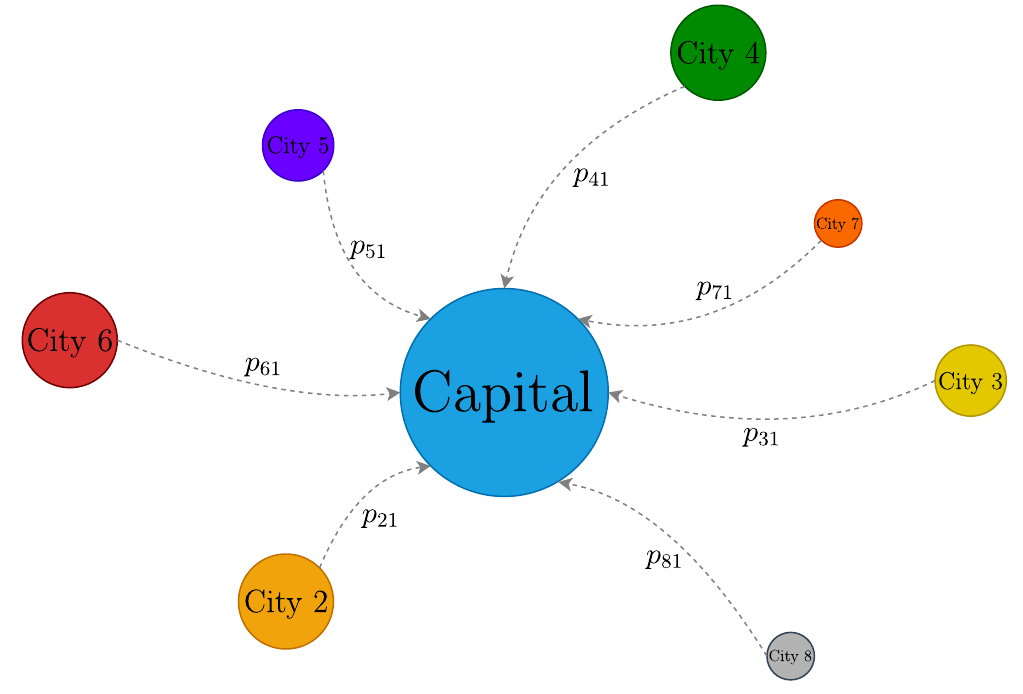}
    \caption{
    {\bf Version of \autoref{fig:cities} considering a metropolitan area and \autoref{assumption:metropolitan_area}.}
    }\label{fig:metropolitean_model}
\end{figure}

From now on, unless indicated, we make the following assumption, which summarizes the idea that commuting majorly occurs between the capital and the other cities as illustrated by~\autoref{fig:metropolitean_model}.

\begin{assumption}[Metropolitan area structure]\label{assumption:metropolitan_area}
    In a metropolitan area, there is a capital, which is a big city with a higher population density and a larger number of inhabitants, along with other connected cities.
    We assume that individuals residing in the capital city both stay and work there, while people in other cities either commute to the capital or work in their home city.
    We further enumerate the cities according to their populations in decreasing order of population size.
    Consequently, matrix $P$ reduces to the form:
    \[
    P = \begin{bmatrix}
        1 & 0 & 0 & \cdots & 0 \\
        p_{21} & p_{22} & 0 & \cdots & 0 \\
        p_{31} & 0 & p_{33} & \cdots & 0 \\
        \vdots & \vdots & \vdots & \ddots & \vdots \\
        p_{K1} & 0 & 0 & \cdots & p_{KK}
    \end{bmatrix}.
    \]
\end{assumption}

\begin{figure}[!htbp]
    \centering
    \includegraphics[width=0.65\textwidth]{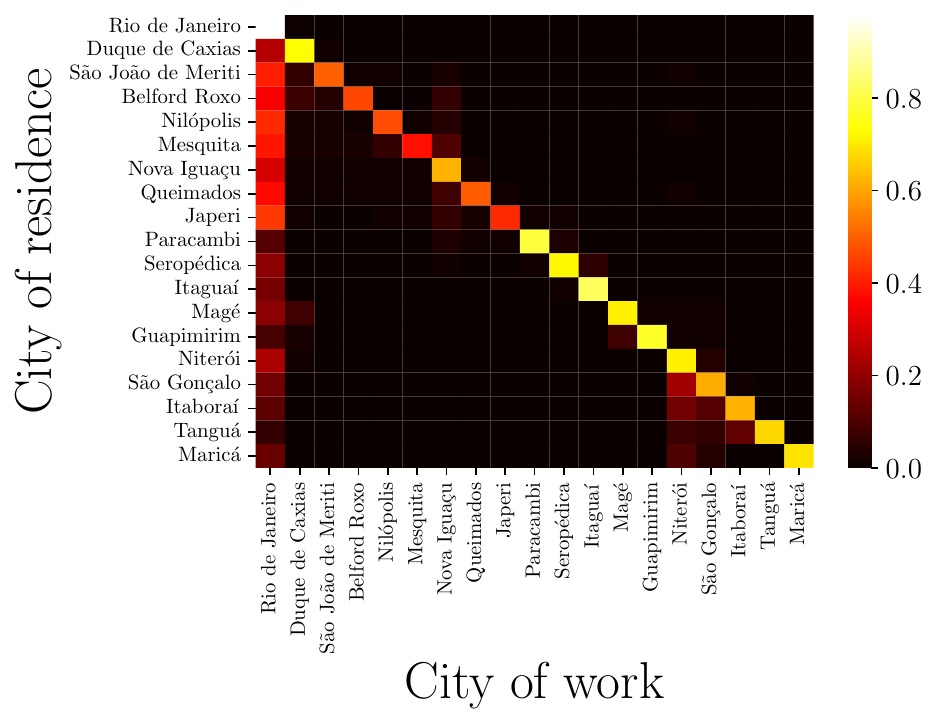}
        \caption{\label{fig:rio_de_janeiro_pmatrix}\textbf{Transition matrix of the Rio de Janeiro metropolitan area.} \\
        A heatmap showing the matrix $P$ in the case of the Rio de Janeiro metropolitan area (mobility data taken from~\cite{sebrae2013}).
        }
\end{figure}

The above assumption adjusts to the reality of several regions. 
For instance, \autoref{fig:rio_de_janeiro_pmatrix} shows the matrix $P$ of the Rio de Janeiro metropolitan area through a heatmap. 
A similar mobility structure is present in the Buenos Aires metropolitan region \cite{gutierrez2021enmodo}. 
Other patterns of mobility may occur in different regions. 
For instance, in Île-de-France, Paris metropolitan area, a portion of individuals from the capital work in the surroundings~\cite{de2021mobilite}.

When applying the specific structure of \autoref{assumption:metropolitan_area} to the dynamics described in equation~\eqref{eq:robot-dance-model}, we derive a different inequality for the reproductive number $\rzero$ (the proof is in \autoref{sec:appendix-proof}).

\begin{theorem}[Inequalities for $\rzero$ under \autoref{assumption:metropolitan_area}]\label{thm:bounds_rzero_metropolitan}
    The basic reproduction number $\rzero$ for system~\eqref{eq:robot-dance-model} under \autoref{assumption:metropolitan_area}, satisfies the following inequalities
    \begin{equation}\label{eq:inequality_rzero_metropolitan}
        \frac{\xi}{\gamma + \mu} \le \rzero \le \frac{1}{\gamma + \mu}\left(\xi + (1-\alpha)\frac{\beta_1}{P_1^{\mathrm{eff}}} \sum_{i=1}^K n_i p_{i1}^2 \right),
    \end{equation}
    where $ \displaystyle \xi = \max_{1\leq i \leq K} \alpha \beta_i + (1-\alpha)p_{ii}\beta_i(1-\delta_{i1})$, where $\delta_{ij}$ is the Kronecker function.
\end{theorem}

Under \autoref{assumption:metropolitan_area}, calculating $\rzero$ reduces to a problem of finding the spectral radius of a diagonal plus a rank-one matrix. 
As far as we know, there is no closed-form expression for it, and obtaining tighter bounds is an open question in the field of {\em Matrix Perturbation Theory\/}~\cite{li2006matrix}.

\begin{remark}
    Higher values of $\gamma$ correspond to faster recovery rates and lower values of $\rzero$.
    As $\alpha$ balances day and night periods, when it approaches $1$, the night period becomes more relevant, and $\rzero$ tends to the maximum of $\rzero^i$, for $i=1,\dots, K$, meaning that infections in the residence city governed the dynamics.
    The parameters $\beta_i$ directly amplify $\rzero$.
    Through numerical experiments, we observed that the highest value among the $\beta_i$ parameters is the most significant driver (see {\em e.g.} \autoref{fig:r0_bounds}). 
    Finally, although the individual population sizes $n_i$ do not directly influence eigenvalue behaviors, the ratios $n_i/n_1$ offer a subdued effect, predominantly guided by $ \displaystyle(1-\alpha)\max_{1\leq i \leq K} \beta_i$.
\end{remark}

\subsection{Vaccination and optimal control problem}

We implement vaccination as a control policy by introducing a time-dependent vaccination rate $u_i$ in each city $i$.
We consider that susceptible individuals can receive the vaccine and immediately move to compartment $R_i$.
Once vaccinated, these individuals no longer contribute to the spread of the disease during the period under consideration.
We could also suppose other sets of hypotheses or model formulations, as we did in \cite{Moschen2023}.
The introduction of vaccination leads to a modified SIR epidemiological model, to which we add a vaccine counter $V_i,$ resulting in the following set of equations: 
\begin{equation}
    \label{eq:robot-dance-model-vaccination}
    \begin{split}
        \frac{dS_i}{dt} &= \mu -\alpha \beta_i S_i I_i - (1-\alpha) S_i \sum_{j=1}^K \beta_j p_{ij} I_j^{\mathrm{eff}} - u_i S_i - \mu S_i, \\
        \frac{dI_i}{dt} &= \alpha \beta_i S_i I_i + (1-\alpha) S_i \sum_{j=1}^K \beta_j p_{ij} I_j^{\mathrm{eff}} - \gamma I_i - \mu I_i, \\
        \frac{dR_i}{dt} &= u_i S_i + \gamma I_i  - \mu R_i, \\
        \frac{dV_i}{dt} &= u_i S_i .
    \end{split}
\end{equation}
This framework assumes that infected individuals are recognizable and do not receive the vaccine. 
This assumption can be challenging in real-life scenarios, but we believe that vaccinating infectious and/or recovered individuals would not change the qualitative results of this work. 
In \autoref{sec:extensions} we discuss a variant of this model, considering vaccination in the workplace.

For the control system \eqref{eq:robot-dance-model-vaccination}, it can be shown that the set  
\[
    \mathcal{C} = \{X \in \R_{\geq 0}^{K \times 4} : X_{i1} + X_{i2} + X_{i3} = 1, \text{ for } i=1,\dots,K\}    
\]
is positively invariant under the flow of system~\eqref{eq:robot-dance-model-vaccination}, for each measurable function $u$. 
For details, see \autoref{sec:appendix-proof}.

Following the steps of \autoref{sec:appendix-r0}, we can calculate the basic reproduction number $\rzero^{\text{vac}}$ for the model \eqref{eq:robot-dance-model-vaccination} with constant vaccination.
Although $\rzero^{\text{vac}}$ does not have a closed-form expression, the vaccination rate inversely impacts $\rzero^{\text{vac}}$.
The inequality presented in equation~\eqref{eq:rzero-inequality} of \autoref{thm:bounds_rzero_generalP} can be adapted for this context by replacing $\rzero^i$ with a modified version $\rzero^{\text{vac}, i}$, where
\[
\rzero^{\text{vac}, i} =  \frac{\mu}{\mu + u_i} \rzero^i = \frac{\mu}{\mu + u_i}\frac{\beta_i}{\gamma + \mu}.
\]

Building upon the formulation of the dynamics, we develop an optimal control problem in which we vaccinate the population of the metropolitan area while considering the cost of vaccination and the cost of hospitalization of infected individuals.
The number of applied doses in city $i$ until time $t$ is
\[
n_i \int_0^t u_i(t)S_i(t) \, dt = n_i V_i(t).
\]
We suppose that a rate $r_h$ of infected individuals are hospitalized with a daily unity cost of $c_h$, and we assume that the unity cost of vaccination is $c_v$.
Therefore, we define the cost functional 
\begin{equation}\label{eq:cost_function}
    J[u_1, \dots, u_K] \coloneqq \sum_{i=1}^K c_v n_i \int_0^T u_i(t) S_i(t) \, dt + c_h r_h n_i \int_0^T I_i(t) \, dt. 
\end{equation}
Due to limitations of resources of the health system, a limited number of people can be vaccinated at each time $t$, so we impose the {\em(mixed control-state)} constraint on the vaccination rate
\begin{equation}\label{eq:mixed_constraints}
    u_i(t)S_i(t) \le \Di, \quad \text{for } t \in [0,T] \text{ and } i=1,\dots,K,
\end{equation}
which implies the following constraint on the daily cap 
\[
\int_{d}^{d+1} u_i(t)S_i(t) \, dt \le \Di, \qquad \text{for } d=0, \dots, T-1.
\]
Through the latter constraint, we indirectly ensure that a proportion of at most $q$ individuals can be immunized by setting 
\[
\sum_{i=1}^K n_i \Di \le q\sum_{i=1}^K n_i,
\]
for some $q \in (0,1).$

\begin{figure}[!htbp]
    \centering
    \includegraphics[width=0.6\textwidth]{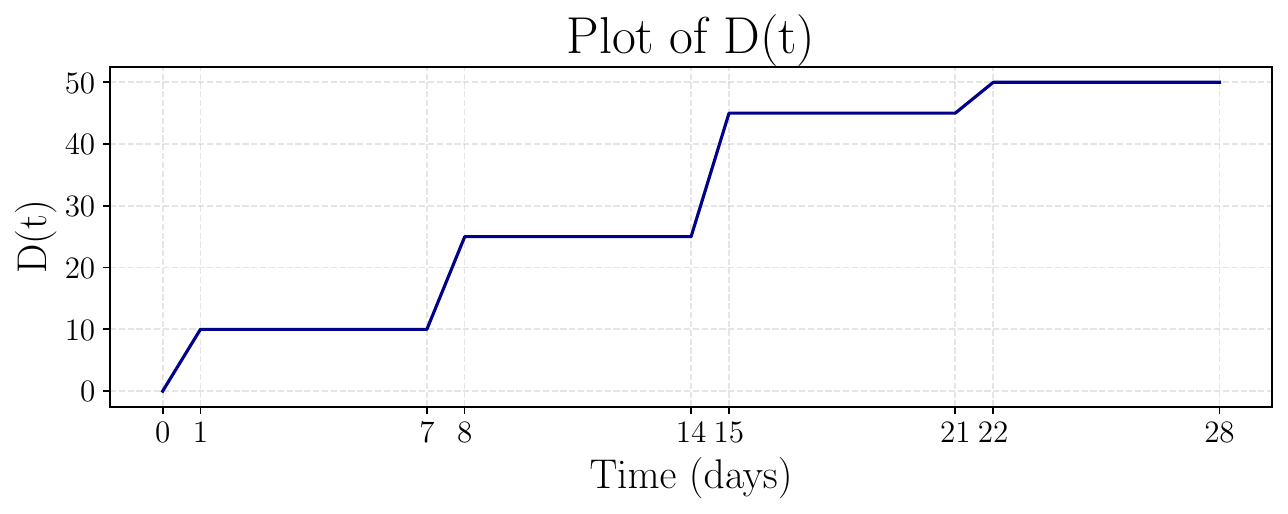}
        \caption{\label{fig:plot_D}\textbf{Example of the function of delivered vaccines.}}
\end{figure}

We consider a scenario where cities receive weekly vaccine shipments from a centralized authority. We use $V_w^{\max}$ to denote the number of vaccines to be received at week $w=0,\dots, T/7 -1$.
From this, we define the function of cumulative vaccine shipments by 
\[
D(t) := \begin{cases}
    \displaystyle\sum_{n=0}^{w-1} V_n^{\max} + V_w^{\max} (t-7w), &t \in [7w, 7w + 1], \\
    \displaystyle\sum_{n=0}^{w} V_n^{\max}, &t \in [7w+1, 7(w+1)],
\end{cases}
\]
for $w = 0, 1, \dots, T/7-1$ and $T$ a multiple of $7$ for simplicity.
The function $D$ is piecewise linear non-decreasing.
Note that, to account for the delay in delivering vaccines across cities and healthcare centers, $D$ increases linearly on the first day of each week, after which it remains constant for the rest of the week.
This induces the following constraint: 
\begin{equation}\label{eq:weekly_shipment}
    \sum_{i=1}^K n_i \int_0^t u_i(t) S_i(t) \, dt = \sum_{i=1}^K n_iV_i(t) \le D(t).
\end{equation}
Observe that any unused vaccines can be employed over the subsequent weeks.
\autoref{fig:plot_D} illustrates an example of this function.

Aggregating the dynamics in~\eqref{eq:robot-dance-model-vaccination}, the cost function in~\eqref{eq:cost_function}, the constraints in~\eqref{eq:mixed_constraints},~\eqref{eq:weekly_shipment} and $u_i \ge 0$, we get the following optimal control problem with $u$ taken in the set of measurable functions from $[0,T]$ to $\R^K$,
\begin{equation}\label{eq:optimal_control_problem}
    \begin{split}
        \min \quad&\sum_{i=1}^K c_v n_i \int_0^T u_i(t) S_i(t) \, dt + c_h r_h n_i \int_0^T I_i(t) \, dt, \\
        \text{s.t. } \quad &\dot{S}_i = \mu - \alpha \beta_i S_i I_i - (1-\alpha) S_i \sum_{j=1}^K \beta_j p_{ij} I_j^{\mathrm{eff}} - u_i S_i - \mu S_i \\
        &\dot{I}_i = \alpha \beta_i S_i I_i + (1-\alpha) S_i \sum_{j=1}^K \beta_j p_{ij} I_j^{\mathrm{eff}} - \gamma I_i - \mu I_i \\
        &\dot{R}_i = u_iS_i + \gamma I_i - \mu R_i\\
        &\dot{V}_i = u_i S_i\\
        &\sum_{i=1}^K n_i V_i(t) \le D(t), \,\, \almost t \in [0,T] \\[0.5px]
        &u_i(t) S_i(t) \le \Di, \,\, \almost t \in [0,T] \\[0.5px]
        &u_i(t) \ge 0,\,\, \almost t \in [0,T] \\[0.5px]
        &S_i(0) = s_{i0},\, I_i(0) = i_{i0},\, R_i(0) = r_{i0},\, V_i(0) = 0.
    \end{split}
\end{equation}

\begin{remark}
    In most of the studies involving vaccination and optimal control, a quadratic cost on the control is considered. 
    That assumption guarantees the existence of a solution, stability of numerical methods, and direct application of Pontryagin's Maximum Principle to get a feedback expression for the optimal control. 
    Nevertheless, we believe that a linear cost in this context is a more realistic representation of operational costs, since it computes the total amount of vaccines or hospitalizations, for instance. 
    In particular, the cost of vaccination should be linear or concave with respect to the quantity of vaccines; a strongly convex assumption does not accurately mirror practical aspects.
\end{remark}

\section{Results}

The core objective of our study is to analyze the effects of different vaccination strategies on the spread of infectious diseases in a metropolitan region modeled by  model~\eqref{eq:robot-dance-model-vaccination}. 
We have carried out a series of experiments that provide a practical platform to evaluate the efficacy of these strategies under controlled conditions.
We start by presenting simulations that delineate the behavior of the basic reproduction number.
This is followed by exploring the impact of constant vaccination rates on the disease dynamics.
We then optimize vaccination strategies, by solving the optimal control problem~\eqref{eq:optimal_control_problem}. 
Lastly, we assess the performance of a practical feedback solution, which we develop by considering a modified problem.

For solving differential equations, we use the explicit Runge-Kutta method of order 5(4) (`rk45'), using the SciPy Python library~\cite{virtanen2020scipy}.
For the optimal control solution, we adopt a {\em first discretize, then optimize\/} approach facilitated by the Gekko Python library~\cite{beal2018gekko}, which uses IPOPT for non-linear optimization.
All experiments were performed on a Linux PC equipped with an AMD Ryzen 9 5950X processor (16 cores) and 128 GB of memory.
The computer code to reproduce the experiments and implement the proposed models is available under a license at \url{https://github.com/lucasmoschen/network-controllability}.

\subsection{Simulations for the basic reproduction number}

Before we introduce vaccination, we examine the impact of the parameters on the basic reproduction number $\rzero$, as described in \nameref{sec:methods} and the bounds for $\rzero$ provided by the aforementioned Theorems~\ref{thm:bounds_rzero_generalP} and~\ref{thm:bounds_rzero_metropolitan}.
For all the simulations throughout this article, we fix $\gamma$ to $1/7$, which means that the recovery time from infection is on average $7$ days, and $\mu = 3.6 \cdot 10^{-5}$ to represent approximately the average life expectancy of $75$ years.

\begin{figure}[!htbp]
    \centering
    \includegraphics[width=0.9\linewidth]{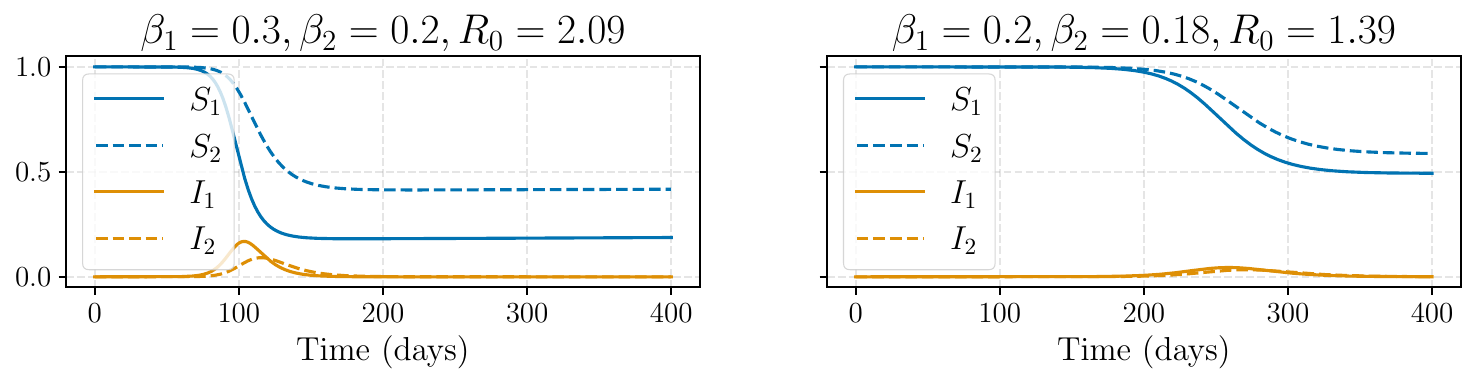}
    \caption{\label{fig:epidemic_behaviour_two_cities}
      \textbf{Dynamics of the Epidemic in Two Cities.} \\ 
      Epidemic dynamics in two cities under two different scenarios. 
      The settings are $\alpha=0.64, p_{21} = 0.2$ and $n_1 = 10$.
      The proportion of susceptible individuals is plotted in blue, and the proportion of infectious individuals in orange.
      The larger city is represented by a solid line, while the smaller city is represented by a dashed line. 
    }
\end{figure}

To open the discussion, we analyze the simpler situation of two cities, where the parameter $p_{21}$, the proportion of individuals that reside in city $2$ and work in city $1$, uniquely determines the transition matrix $P$.
Let us set $n_2=1$ since only the ratio $n_i/n_1$ is relevant to $\rzero$, not the magnitude.
\autoref{fig:epidemic_behaviour_two_cities} presents the dynamics of the epidemic in two different scenarios for $\beta,$ considering a single infected initial individual in the population of the first city.
Notice that higher values of $\rzero$ give higher and faster epidemic peaks.

\begin{figure}[!htbp]
    \centering
    \includegraphics[width=\textwidth]{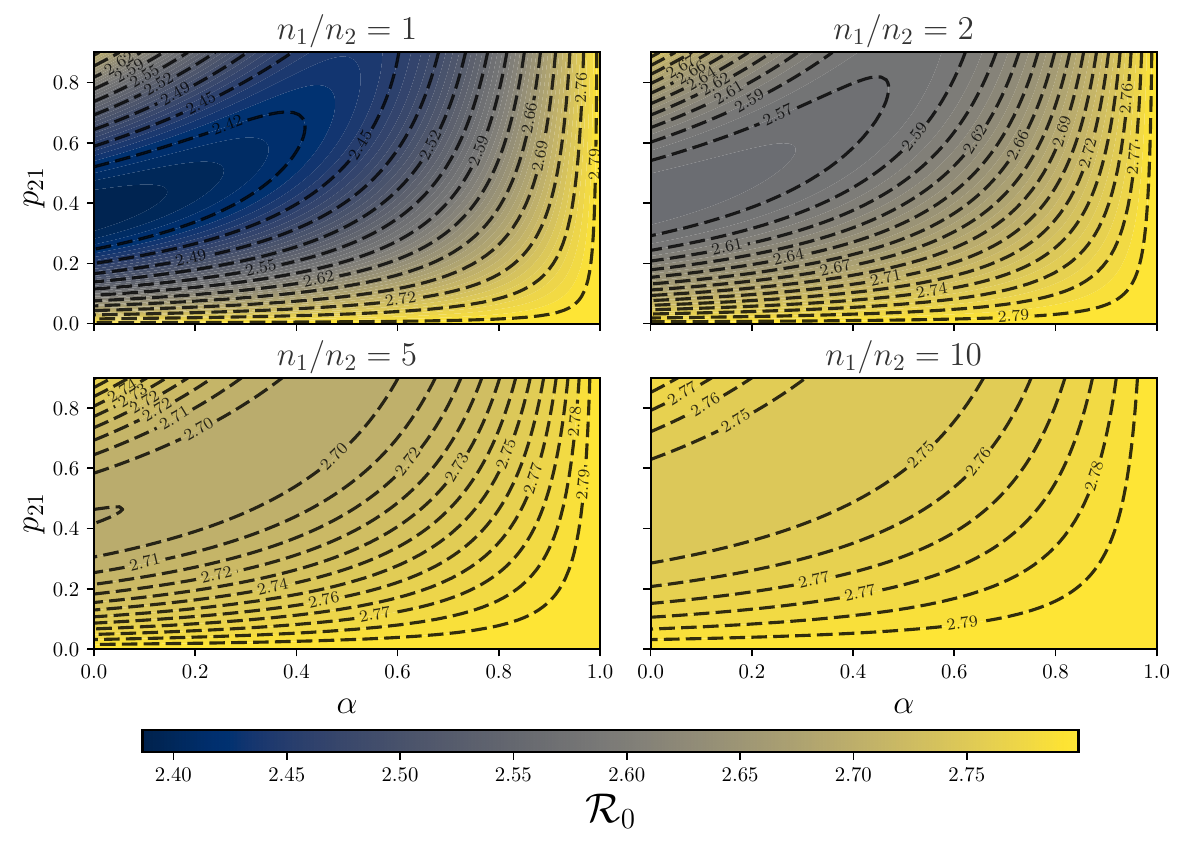}
    \caption{\label{fig:r0_function_alpha_p21}\textbf{Basic reproduction number as function of $\alpha$ and $p_{21}$ for two cities.} \\
    Contour plot of the values of $\rzero$ for different values of $\alpha \in (0,1)$, $p_{21} \in (0,0.9)$ and $n_1/n_2 = (1,2,5,10)$  considering $\beta = (0.4, 0.1)$.
    }
\end{figure}

In \autoref{fig:r0_function_alpha_p21} we show the level curves of $\rzero$ when $p_{21},$ $n_1/n_2$ and $\alpha$ vary.
For the chosen parameters $\beta, \gamma$ and $\mu$, we have $\rzero^1 > 1$ and $\rzero^2 < 1,$ which represents a situation for which \autoref{cor:lower_upper_bound} cannot determine whether $\rzero$ is greater or smaller than 1.
Looking at \autoref{fig:r0_function_alpha_p21}, we notice that higher values of $\alpha$ increase the value of $\rzero$.
The impact of $p_{21}$ is nonlinear and complex, but its behavior is similar to a quadratic function in $\rzero$.
On the other hand, the impact of $p_{21}$ and $\alpha$ is small when the population proportion $n_1/n_2$ increases. 
Additionally, see the histogram of \autoref{fig:r0_function_alpha_p21_smaller_beta}, which illustrates the situation where $\rzero^1$ is close to 1.

\begin{figure}[!htbp]
  \centering
  \includegraphics[width=\textwidth]{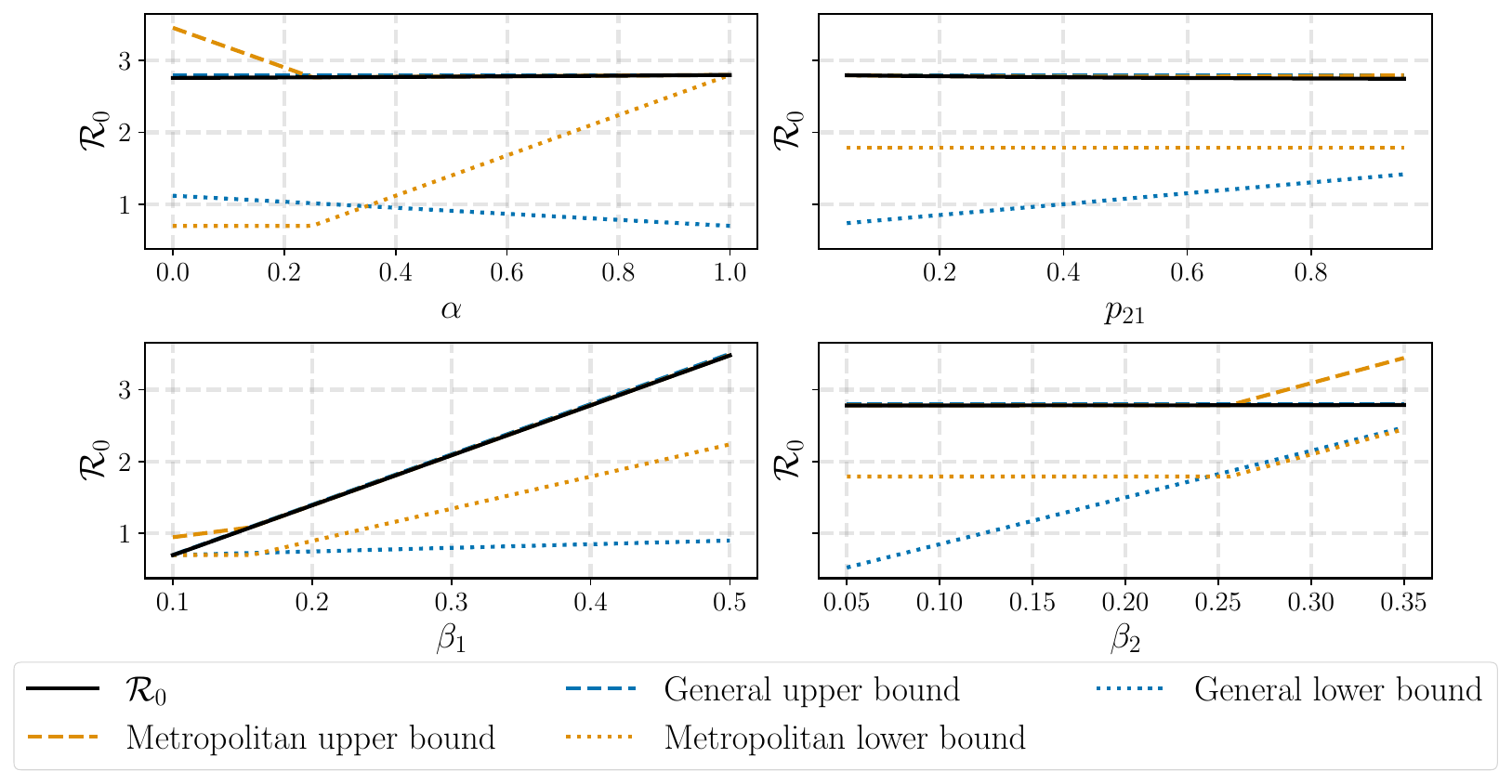}
  \caption{\label{fig:r0_bounds}\textbf{Impact of the parameters on the bounds of $\rzero$.} \\
  Using the same setting as in \autoref{fig:epidemic_behaviour_two_cities}, we visualize the impact of $\alpha, p_{21},\beta_1$ and $\beta_2$ on $\rzero$ (depicted in black).
  In orange, we see the bounds from \autoref{thm:bounds_rzero_metropolitan}, and in blue, the bounds from \autoref{thm:bounds_rzero_generalP}.
  The dotted lines are the lower bounds while the dashed lines are the upper bounds.}
\end{figure}

In \autoref{fig:r0_bounds}, we analyze the bounds for $\rzero$ provided by \autoref{thm:bounds_rzero_generalP} and \autoref{thm:bounds_rzero_metropolitan}.
We mainly observe that $\rzero$ is primarily driven by $\beta_1$, the highest effective transmission rate, while $\alpha$, $p_{21}$, and $\beta_2$ have little impact.
Higher values of $\alpha$ improve the bounds of \autoref{thm:bounds_rzero_metropolitan}, depicted in orange, mainly due to the term factor of $(1-\alpha)$ appearing on the right-hand side of \eqref{eq:inequality_rzero_metropolitan}.
However, the fourth graph (where $\beta_2$ varies) illustrates the fact that $\rho(\alpha B + (1-\alpha)\Lambda)$ remains constant until $\alpha \beta_2 + (1-\alpha)\beta_2 p_{22} < \alpha \beta_1$, beyond which the upper bound (in orange dashed line) gets worse.

In the context of multiple cities, we consider a scenario of $K=5$ cities.
We first examine five distinct mobility scenarios: (I) metropolitan area where $10\%$ of the residents of each city work in the capital; (II) metropolitan area where $40\%$ of each city works in the capital; (III) all cities are interconnected, where 60\% of the residents work in their city, while the remaining are distributed evenly across the other four cities; (IV) a variant of the metropolitan area where 30\% of each city's residents work in the capital and 10\% of the capital's workforce is employed in the other cities; (V) no commuting between cities.

\begin{figure}[!htbp]
    \centering
    \includegraphics[width=\textwidth]{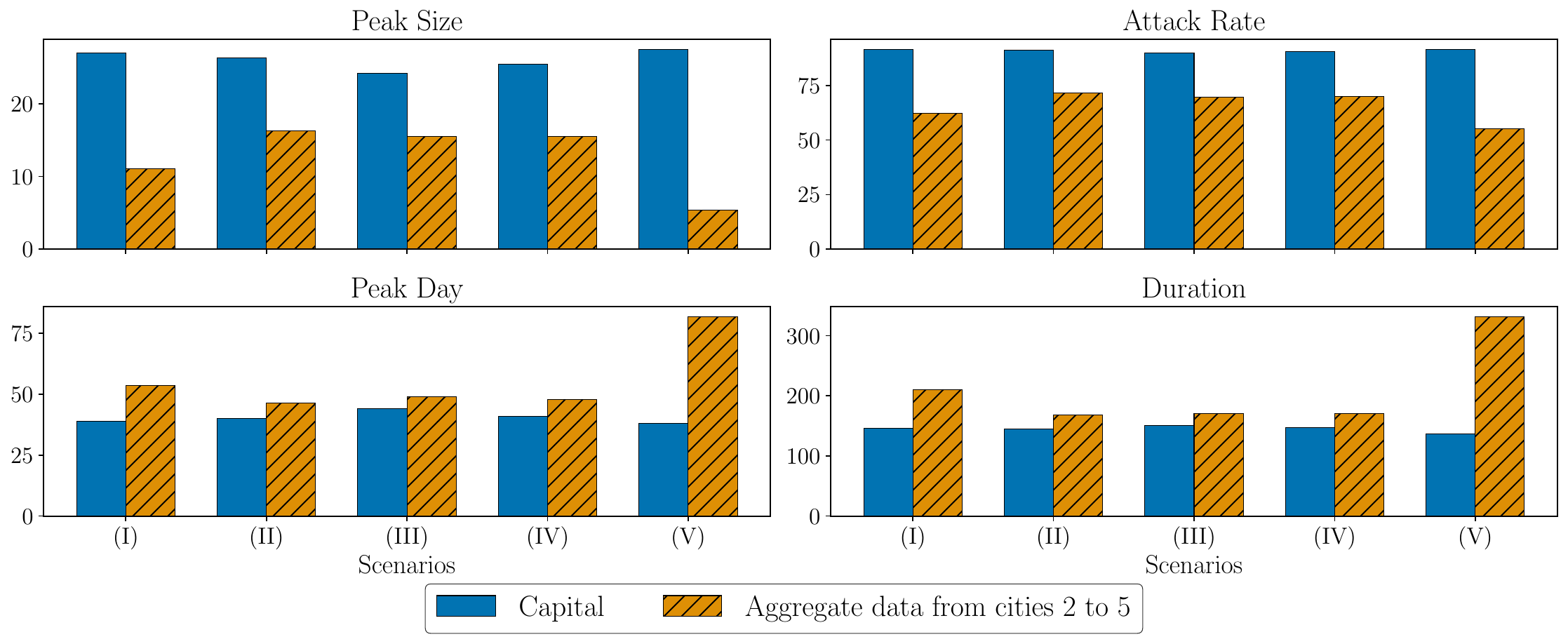}
    \caption{\label{fig:summary_scenarios_matrix_P}\textbf{Simulation results for different structures of the transition matrix.} \\
    Comparison of Peak Size (maximum proportion of infectious individuals), Peak Day (day of the peak size), Duration (time from epidemic onset to the day when the proportion of infectious individuals achieves $10^{-5}$), and Attack Rate (proportion of individuals who contract the disease during $350$ days of the epidemic) across different mobility scenarios.
    The aggregated data is the weighted average, considering their population size, from cities $2$ to $5$.
    For this experiment, $\beta = (0.4, 0.25, 0.2, 0.15, 0.1)$, $\alpha = 0.64$, and the population sizes are $10^5\cdot(50, 10, 10, 1, 1)$.
    Lastly, the initial conditions are $I_1(0) = I_2(0) = I_3(0) = 10^{-4}$ and no recovered individuals.
    }
\end{figure}

Our findings, summarized in \autoref{table:summary_scenarios_matrix_P} and illustrated in \autoref{fig:summary_scenarios_matrix_P}, 
reveal that in the capital city -- compared to the other cities -- the peak size is higher, the peak day occurs earlier, the outbreak (period between $t=0$ and the time at which the proportion of infectious individuals falls below $10^{-5}$) lasts less and the attack rate is larger.
Additionally, higher commuting rates correlate with a shorter epidemic span, as scenarios (II) and (III) exhibit and the capital's peak day tends to dictate the overall trajectory of the disease spread across the network.
Further, the outcomes of scenarios (II) and (III) are similar, which indicates that depending on the structure of the network, we may approximate it using the metropolitan hypothesis in \autoref{assumption:metropolitan_area}.

\begin{figure}[!htbp]
    \centering
    \includegraphics[width=\textwidth]{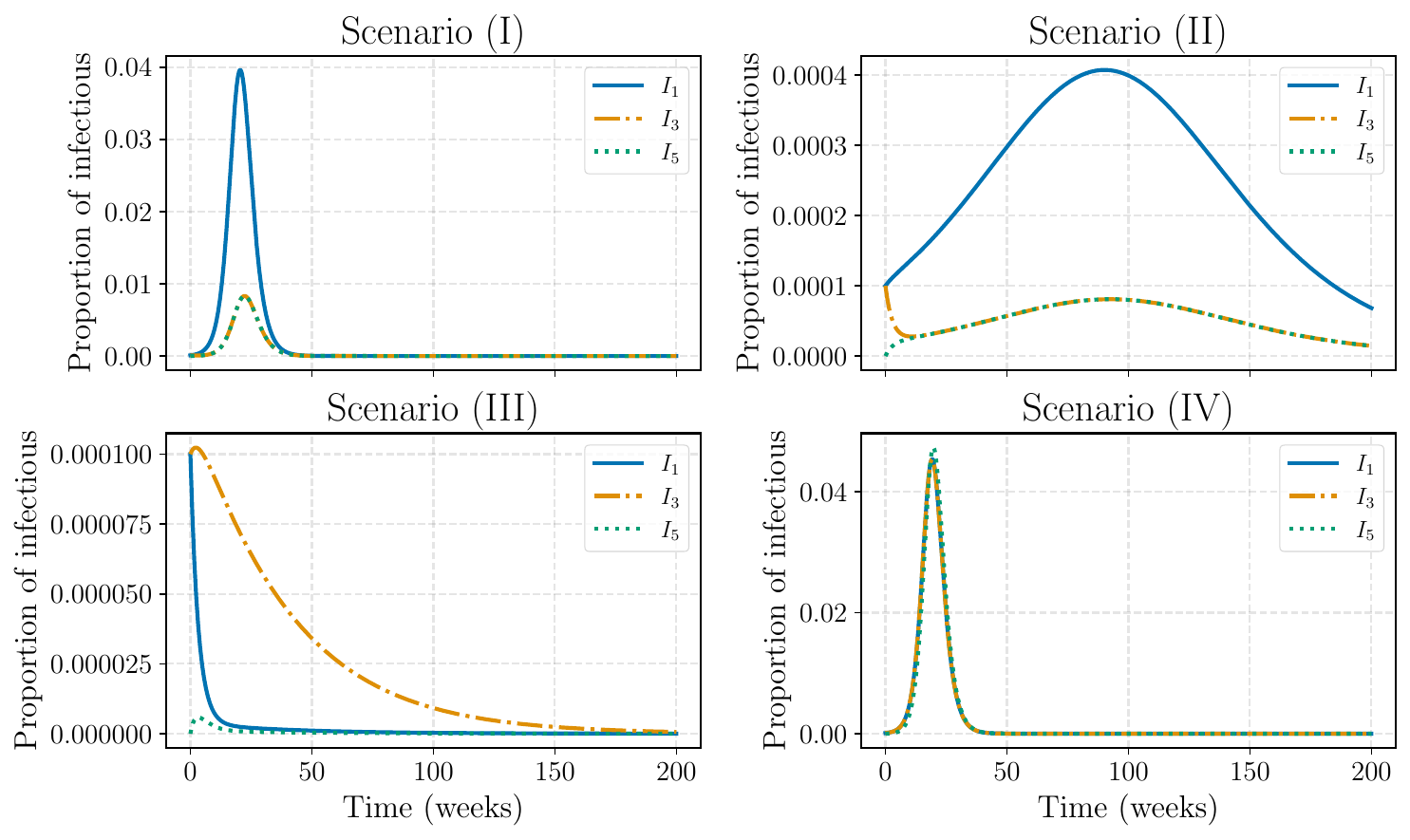}
\caption{\label{fig:dynamics_infectious_different_beta_values}\textbf{Infectious curves for different scenarios of $\beta$.} \\
Curves of infectious individuals from cities $1$, $3$, and $5$ under four scenarios of the $\beta$ specification. 
Here $\alpha = 0.64, p_{i1} = 0.2$ for $i \ge 2$ and $n_1 = 5n_2 = 5n_3 = 50n_4 = 50n_5$ $= 5\cdot 10^6$.
The four scenarios considered are (I) $\beta = (0.2,0.1,0.1,0.1,0.1)$, indicating $\rzero^1 > 1$ and $\rzero^i < 1$ for $i\neq 1$; (II) $\beta = (0.15,0.1,0.1,0.1,0.1)$, a situation similar to (I), but with a slower transmissibility rate, even though $\rzero > 1$; (III) $\beta = (0.1, 0.1, 0.15, 0.1, 0.1)$, a situation similar to (II), but with the highest transmissibility rate not in the capital. 
This scenario results in $\rzero < 1$; (IV) $\beta = (0.2, 0.2, 0.2, 0.2, 0.2)$, where all cities have the same transmissibility rate.
}
\end{figure}

In \autoref{fig:dynamics_infectious_different_beta_values}, we vary the infection rate $\beta.$ We consider an experiment of four different scenarios in five cities and depict only three cities to better exhibit the results.
When comparing scenarios (I) and (II), both have $\rzero$ greater than 1. 
However, a minor adjustment in $\beta_1$ from $0.2$ to $0.15$ results in a nearly $100$-fold decrease in peak size, changing the entire aspect of the epidemic.
Interestingly, we observed that even if $\rzero^i < 1,$ for $i = 2, \ldots, 5$, the epidemic still manages to spread in these cities.
Turning our attention to scenario (III), this setting represents a situation where $\rzero < 1$, achieved by modifying the city with the highest transmissibility rate: from the capital to the city $3$, the second largest city. 
This implies that the capital, being the main hub of mobility, plays a significant role in accelerating the spread of the disease.
Finally, scenario (IV) fixes the same infection rate for all cities. 
Notice in this case that the behavior of the epidemic is very similar for all cities and mobility does not play a big role in the capital.

\subsection{Constant vaccination rates}

We now turn our attention to the analysis of the vaccination policy considering a constant rate $u_i$ for each city $i$, as outlined by the equations in~\eqref{eq:robot-dance-model-vaccination}.

\begin{figure}[!htbp]
    \centering
    \includegraphics[width=0.7\textwidth]{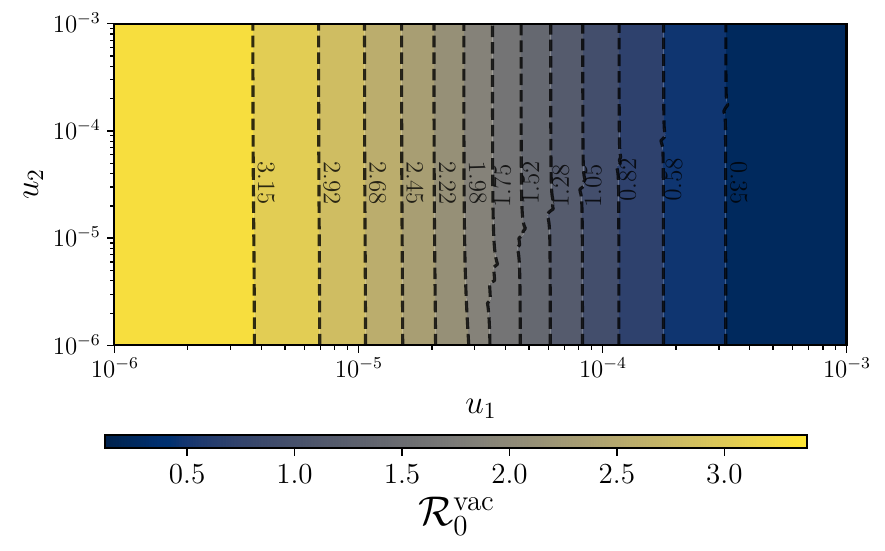}
    \caption{\label{fig:r0_function_vaccination}\textbf{Basic reproduction number $\rzero^{\mathrm{vac}}$ in the vaccination model.} \\
    The basic reproduction number $\rzero^{\mathrm{vac}}$ for the model described in~\eqref{eq:robot-dance-model-vaccination} as a function of $u_1$ and $u_2$ across $K=2$ cities.
    The settings are: $\beta = (0.5, 0.3)$, $\alpha = 0.64$, $p_{21} = 0.2$ and the ratio $n_1/n_2 = 10$.
    Both axes $u_1$ and $u_2$ are in the logarithmic scale.
    }
\end{figure}

In this scenario, we consider an ongoing epidemic that is subject to a vaccination strategy to control it, starting on a specified day $s$.
In the two-city scenario, we set $\beta = (0.5, 0.3)$ to ensure that $\rzero^1 > 1$ and $\rzero^2 > 1$.
The vaccination rates are varied within the range $u = (u_1,u_2) \in {[10^{-6}, 10^{-3}]}^2$ to generate \autoref{fig:r0_function_vaccination}, which is presented on the $\log$-$\log$ scale.
For this setting, the basic reproduction number without vaccination is $\rzero \approx 3.48$.
Numerically, by a fitting method, we observe that
\begin{equation}\label{eq:r0_vaccine_formula}
    \rzero^{\mathrm{vac}}(u_1, u_2) \approx \frac{\rzero}{1 + u_1 \cdot b(u_2)},
\end{equation}
where $b$ is approximately a constant function of $u_2$, as we observe in \autoref{fig:r0_function_vaccination}.
By this figure and by equation \eqref{eq:r0_vaccine_formula}, we infer that the vaccination rate of the capital city governs $\rzero^{\mathrm{vac}}$.

To analyze the significance of vaccinating the capital, we perform a simplified version of the optimal control problem introduced in equation~\eqref{eq:optimal_control_problem} only considering the constant vaccination strategies. 
We use a constant vector $u \in {[0, 0.05]}^K$ as the decision variable. 
In this constant vaccination case, we can simplify the constraints and consider the following:
\[
\sum_{k=1}^K V_k(T) n_k \le 0.8\sum_{k=1}^K n_k
\]
to ensure that the total number of vaccinated individuals at the final time $T$ is at most 80\% of the population. 
We set $T = 8 \times 7 = 56$ days, the hospitalization rate $r_h = 0.1$ and the hospitalization cost $c_h = 1000$.
Vaccination starts at the $70$th day after the first infection in the capital.
Finally, we consider three different scenarios for the parameter $\beta$ and three specifications of the unity cost of a vaccine $c_v$.
As we can see in \autoref{tab:vaccination_rates_different_scenarios}, it is evident that in all scenarios, the capital city not only receives the majority of vaccines but also achieves the highest vaccination rate, particularly reaching the upper bound in $\beta^{(2)}$ and $\beta^{(3)}$ scenarios.
In the first scenario, which is characterized by a uniform infection rate across the cities, vaccine distribution is more equitable, yet the capital still maintains an advantage. 
The behavior is similar when $u_j$ is allowed to vary up to $0.1$ for each city $j$.
However, it is important to note that the relevance of vaccinating the capital can diminish in scenarios where the susceptible population decreases rapidly, as vaccine allocation depends significantly on the size of the available population.

\begin{table}[!htbp]
    \centering
    \begin{tabular}{c c c c c c c}
        $\beta$ & $c_v$ & Capital & City 2 & City 3 & City 4 & City 5 \\\toprule{}
        \multirow{3}{*}{$\beta^{(1)}$} & $c^1_v$ & $\boldsymbol{4.26 \cdot 10^{-2}}$ & $3.41 \cdot 10^{-2}$ & $3.41 \cdot 10^{-2}$ & $3.41 \cdot 10^{-2}$ & $3.41 \cdot 10^{-2}$ \\
        & $c^2_v$ & $\boldsymbol{4.26 \cdot 10^{-2}}$ & $3.41 \cdot 10^{-2}$ & $3.41 \cdot 10^{-2}$ & $3.41 \cdot 10^{-2}$ & $3.41 \cdot 10^{-2}$ \\
        & $c^3_v$ & $\boldsymbol{4.26 \cdot 10^{-2}}$ & $3.41 \cdot 10^{-2}$ & $3.41 \cdot 10^{-2}$ & $3.41 \cdot 10^{-2}$ & $3.41 \cdot 10^{-2}$ \\
        \midrule{}
        \multirow{3}{*}{$\beta^{(2)}$} & $c^1_v$ & $\boldsymbol{5.00 \cdot 10^{-2}}$ & $\boldsymbol{5.00 \cdot 10^{-2}}$  & $9.63 \cdot 10^{-3}$ & $9.63 \cdot 10^{-3}$ & $3.42 \cdot 10^{-8}$ \\
        & $c^2_v$ & $\boldsymbol{5.00 \cdot 10^{-2}}$ & $\boldsymbol{5.00 \cdot 10^{-2}}$  & $9.63 \cdot 10^{-3}$ & $9.63 \cdot 10^{-3}$ & $3.02 \cdot 10^{-7}$ \\
        & $c^3_v$ & $\boldsymbol{5.00 \cdot 10^{-2}}$ & $1.38 \cdot 10^{-2}$ & $1.65\cdot 10^{-8}$ & $1.70 \cdot 10^{-7}$ & $1.62 \cdot 10^{-7}$ \\
        \midrule{}
        \multirow{3}{*}{$\beta^{(3)}$} & $c^1_v$ & $\boldsymbol{5.00 \cdot 10^{-2}}$ &$\boldsymbol{5.00 \cdot 10^{-2}}$  & $1.06 \cdot 10^{-2}$ & $1.54 \cdot 10^{-7}$ & $5.01 \cdot 10^{-8}$ \\
        & $c^2_v$ & $\boldsymbol{5.00 \cdot 10^{-2}}$ &$\boldsymbol{5.00 \cdot 10^{-2}}$ & $1.06 \cdot 10^{-2}$ & $1.76 \cdot 10^{-7}$ & $3.36 \cdot 10^{-8}$ \\
        & $c^3_v$ & $\boldsymbol{5.00 \cdot 10^{-2}}$ & $7.65 \cdot 10^{-9}$ & $7.65 \cdot 10^{-9}$ & $7.91 \cdot 10^{-8}$ & $6.20 \cdot 10^{-8}$ \\
        \bottomrule{}
    \end{tabular}
    
    \caption{\label{tab:vaccination_rates_different_scenarios}\textbf{Optimal constant vaccination rates across cities for different parameter settings:} 
    vaccination rates in five different cities under nine parameter settings combining three values for $\beta$ and three for $c_v$. 
    We set $\beta^{(1)} = (0.3, 0.3, 0.3, 0.3, 0.3)$; $\beta^{(2)} = (0.3, 0.2, 0.12, 0.12, 0.1)$; and $\beta^{(3)} = (0.3, 0.12, 0.12, 0.1, 0.05)$, and $c_v^1=0.001$; $c_v^2 = 0.1$; and $c_v^3 = 10$.
    Each row corresponds to a different combination of $\beta$ and $c_v$ values, representing different scenarios of disease transmission rate and vaccination cost. 
    For each row, in bold, we highlight the highest rate.
    }
\end{table}

As seen in \autoref{tab:vaccination_rates_different_scenarios}, when the first day of vaccination is determined by the number of infectious individuals, for instance, the day on which $1\%$ of the total population is infected, all scenarios indicate that the capital should receive vaccines at the highest rate, regardless of $c_v$.
The same result is obtained when the initial susceptible population is uniform across the cities.
This analysis showcases that it is preferable to vaccinate the capital at a higher rate.

\subsection{Performance of time-variable vaccination strategies}

\begin{figure}[!hbtp]
    \centering
    \includegraphics[width=\textwidth]{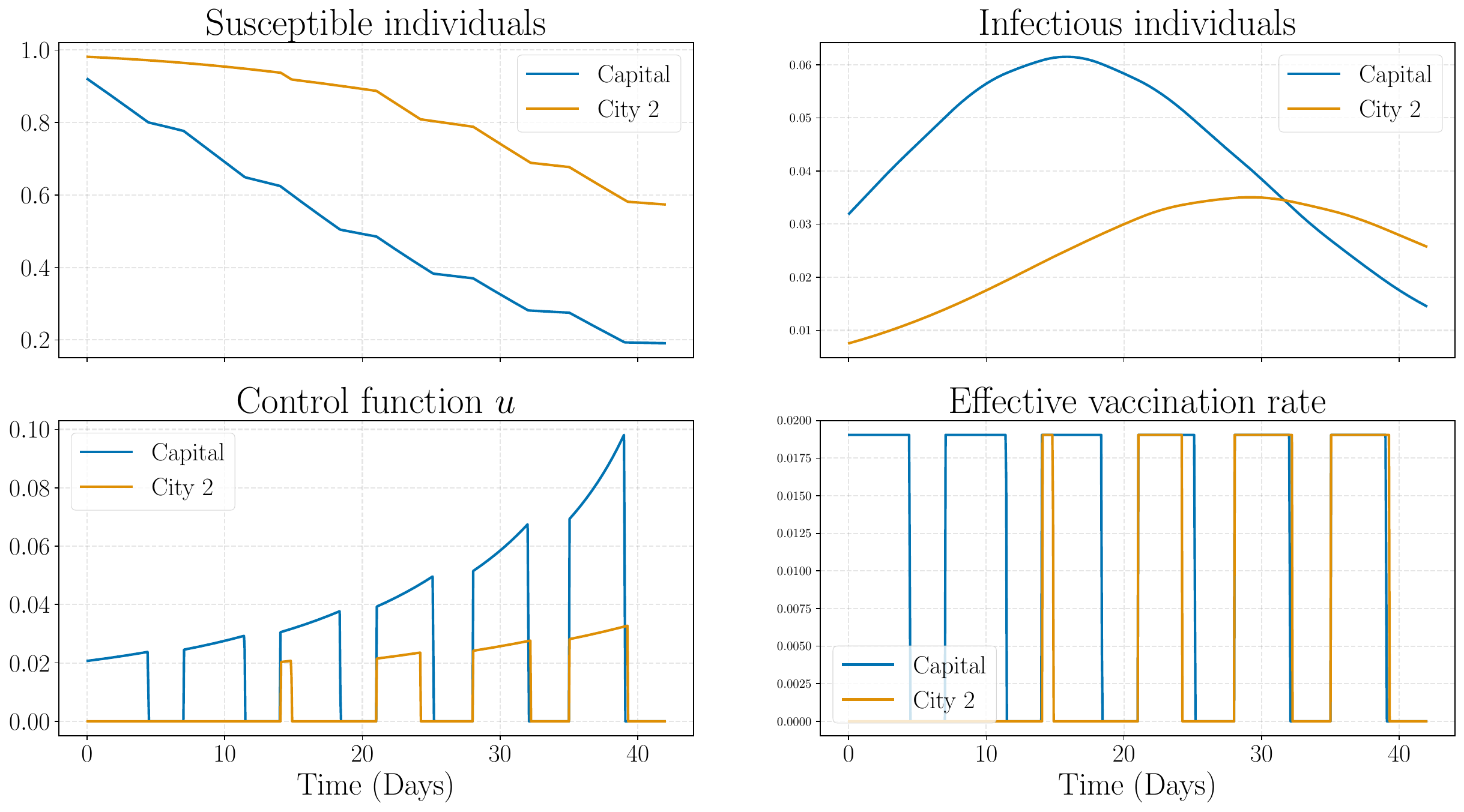}
    \caption{\label{fig:numerical_simulations_control_two_cities}\textbf{Optimal trajectories and control for two-city interaction.} \\ 
    Optimal trajectory of the proportion of susceptible and infectious individuals in two cities, the capital in orange and the second city in blue.
    The fourth subplot illustrates the variables $u_i(t)S_i(t) \in \{0, \Di\}, \, i=1,2$.
    The settings for this experiment are: $\beta = (0.25, 0.18)$, $\alpha = 0.64$, $p_{21} = 0.2$, $n_1 = 10^6$ and $n_2 = 10^5$.
    For the optimal control problem,  $c_v = 0.01$, $c_h = 1000$, $v^{\mathrm{max}}_1 = v^{\mathrm{max}}_2 = 0.8/42$ and a weekly cap allowing the vaccination of at most $1/13$ of the population per week.
    }
\end{figure}

In this section, we perform numerical simulations that incorporate vaccination as a time-variable control function.
By integrating vaccination strategies into our metropolitan area model, the goal is to provide a comprehensive understanding of disease dynamics and control measures. 
We start by examining the interaction between two cities, as shown in \autoref{fig:numerical_simulations_control_two_cities}.
This figure illustrates the optimal trajectory and control for a capital city with a population $10$ times greater than the second city and a higher infection rate. 
Vaccination begins on the $100$th day following the arrival of the first infected individual in the capital.

\begin{figure}[!htbp]
    \centering
    \includegraphics[width=0.8\textwidth]{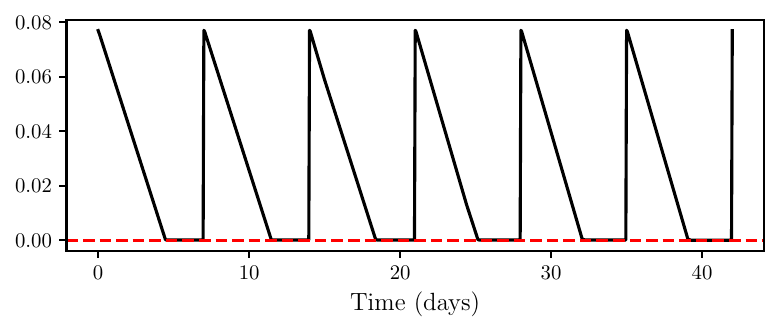}
    \caption{\label{fig:weekly_cap_two_cities}\textbf{Weekly constraint analysis.} \\
    Function $D(t) - \sum_{k=1}^K n_k V_k(t)$ for the experiment presented in \autoref{fig:numerical_simulations_control_two_cities}.
    }
    
\end{figure}

Notably, during the initial two weeks, the optimal solution allocates all vaccines to the capital, administering them at the maximum possible rate according to the weekly cap.
Subsequently, as the susceptible population in the capital decreases to around $60\%$, it becomes advantageous to start vaccinating the second city, which has a lower daily cap.
The fourth graph illustrates the {\em bang-bang}~\cite{schattler2012geometric} behavior of the {\em effective vaccination rate} $u_iS_i,$ this is, in which $u_i^*(t)S_i^*(t) \in \{0, \Di\}$ holds for $i=1,2$, along the whole interval $[0,T].$
The gap of the weekly cap constraint is depicted in \autoref{fig:weekly_cap_two_cities}.
We assign a high unity cost to hospitalization $c_h = 1000$ in contrast to a relatively low unity cost for vaccination, with $c_v = 0.01$ to reflect the burden of hospitalization. 
Both healthcare resources and patient well-being are comparatively way more expensive than the minor inconvenience and price of vaccination. 
Furthermore, the benefits of vaccination are amplified when administered on a large scale.
This simulation took around $33$ seconds to run.

\begin{figure}[!btph]
    \centering
    \includegraphics[width=\textwidth]{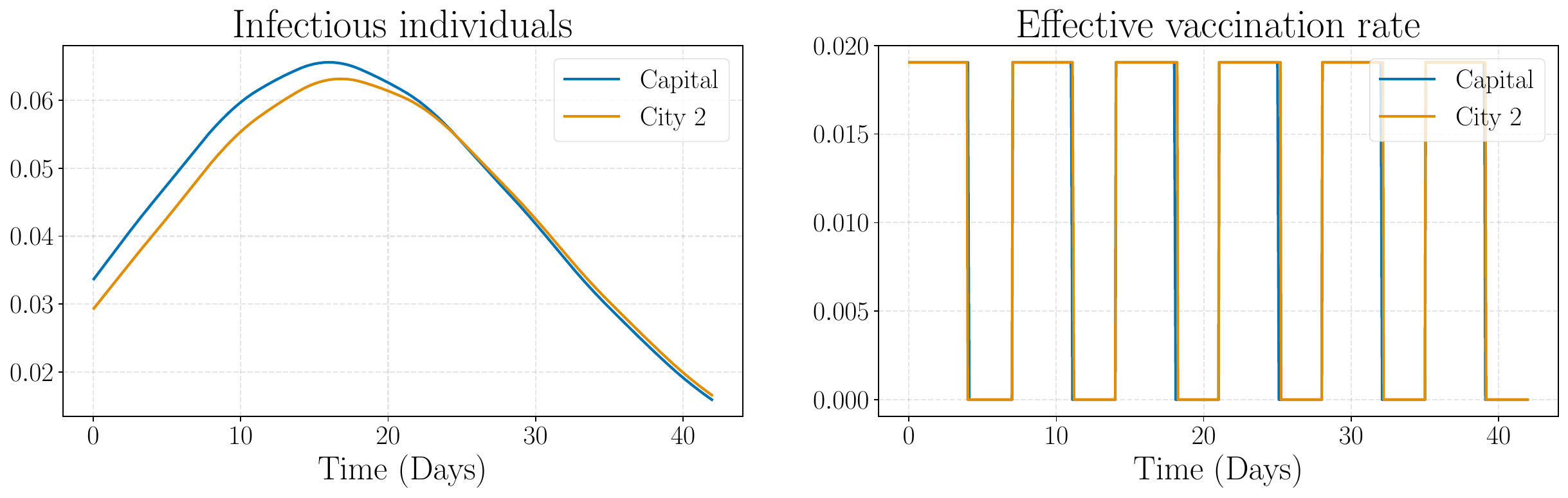}
    \caption{\label{fig:numerical_simulations_control_two_cities_fig3}\textbf{Optimal trajectories and control for two-city interaction for uniform infection rates.} \\
    Optimal trajectory of the proportion of susceptible and infectious individuals in two cities and the optimal control using the same settings as \autoref{fig:numerical_simulations_control_two_cities} except $\beta = (0.25, 0.25)$.
    }
\end{figure}

When we equalize the infection rates of the two cities, setting $\beta_1 = \beta_2 = 0.25$, while keeping the other parameters fixed, we obtain the solution in \autoref{fig:numerical_simulations_control_two_cities_fig3}.
One can see that the behavior of the solution is similar to that of the previous experiment but with a uniform optimal solution across both cities.
This suggests that the transmission rate $\beta$ is the most important driver when deciding vaccination strategies.
Our prior conclusion about the importance of the susceptible population in deciding which city retains the vaccination is also evident in the figure.

\begin{figure}[!hbtp]
    \centering
    \includegraphics[width=\textwidth]{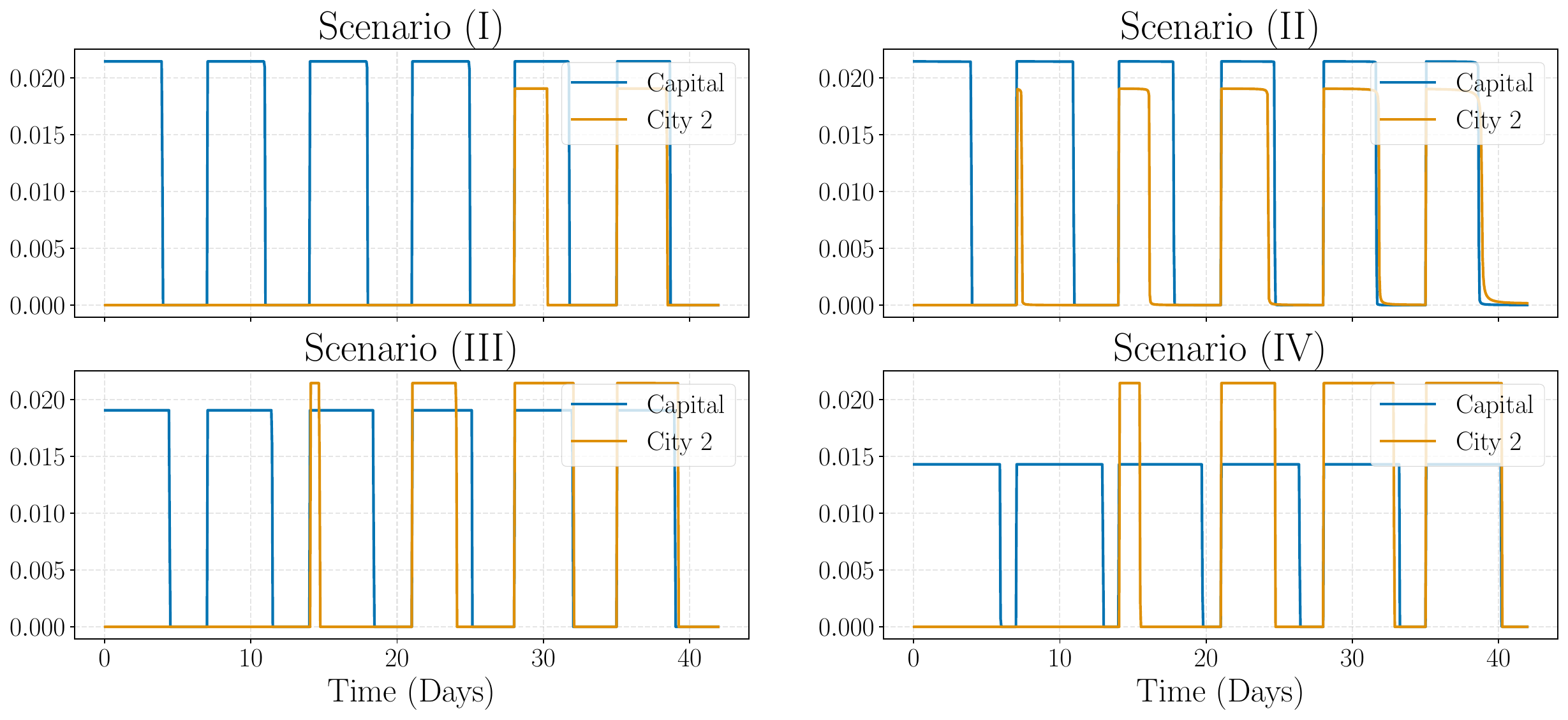}
    \caption{\label{fig:numerical_simulations_control_two_cities_fig2}\textbf{Optimal effective vaccination rate for two-city interaction.} \\
    Scenario (I) considers the same setting as \autoref{fig:numerical_simulations_control_two_cities} except for the first infection happening in the second city and $v_1^{\max} = 0.9/42$ (while $v_2^{\max} = 0.8/42$).
    Scenario (II) considers $I_1(0) = I_2(0) = 0.05$ and $R_1(0) = R_2(0) = 0.02$.
    Scenarios (III) and (IV) fix the same parameters of \autoref{fig:numerical_simulations_control_two_cities} but with different maximum rates $v_1^{\max}$ and $v_2^{\max}$.
    In (III), we set $v_1^{\mathrm{max}} = 0.8/42 < v_2^{\mathrm{max}} = 0.9/42$, while in (IV), $v_1^{\mathrm{max}} = 0.6/42 < v_2^{\mathrm{max}} = 0.9/42$.
    }
\end{figure}

Finally, we consider four different scenarios: first infection in the second city (Scenario (I)), uniform initial conditions for the control problem (Scenario (II)), and two that express higher transmissibility in the capital, but a lower vaccination rate limitation (Scenarios (III) and (IV)).
The result is shown in \autoref{fig:numerical_simulations_control_two_cities_fig2}.
It establishes that the vaccination always starts in the capital in all scenarios, even with the infection starting in the second city. 
Moreover, Scenarios (III) and (IV) confirm that this behavior does not come from the maximum effective vaccination rate, highlighting that even though we can vaccinate more rapidly in the second city, the capital starts first.

Having considered the simplified two-city case, we extend our analysis to more cities in the metropolitan area.
We first observe that the dimensionality of the problem escalates quickly, which makes efficient implementations a challenging aspect yet to be addressed in the literature.
\citeauthor{lemaitre2022optimal} (\citeyear{lemaitre2022optimal})~\cite{lemaitre2022optimal} consider a similar optimal control problem and offer a viable solution, which we could further explore in future research.

\begin{figure}
    \centering
    \includegraphics[width=0.9\textwidth]{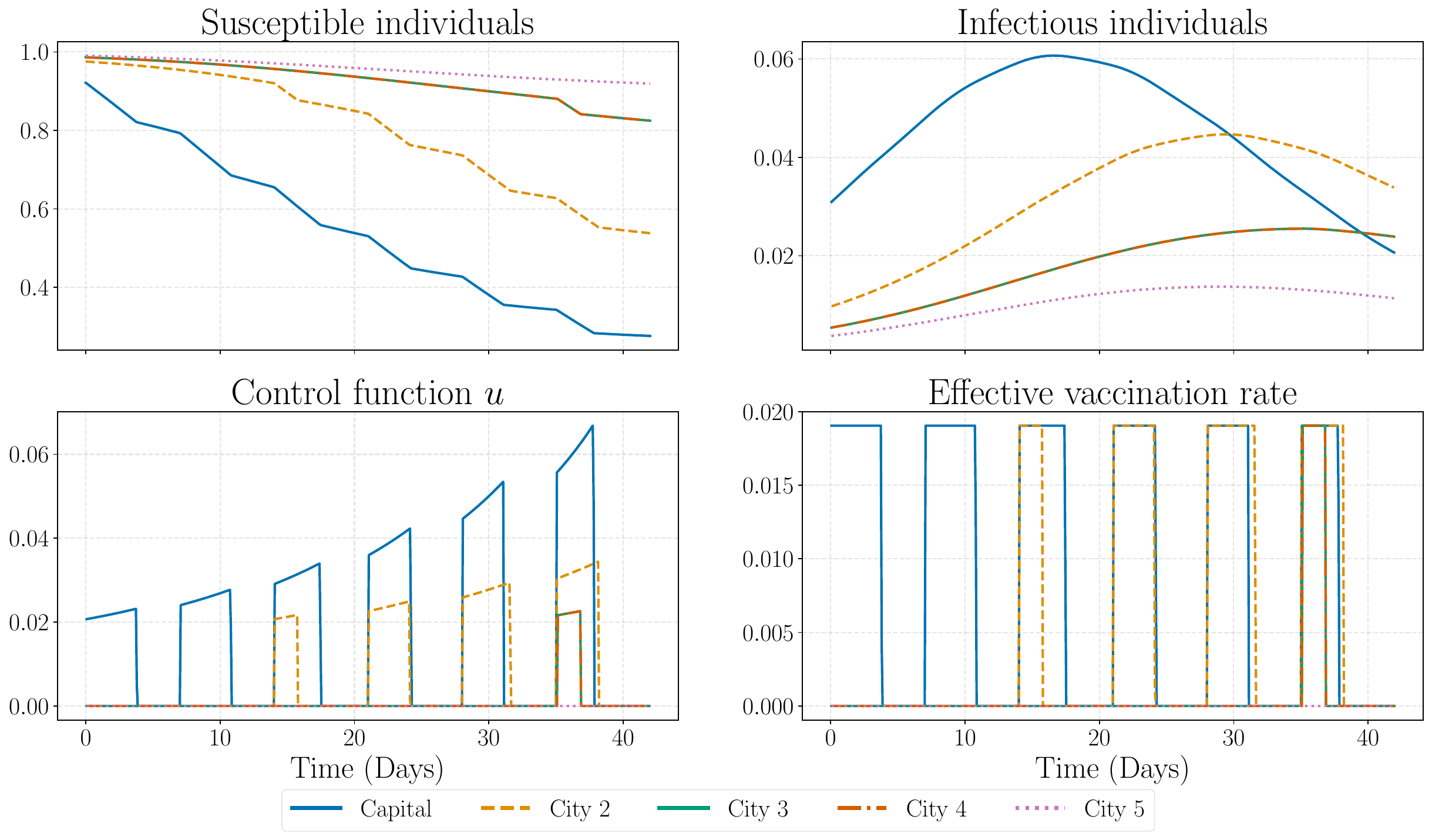}
    \caption{\label{fig:numerical_simulations_control_fig4}\textbf{Optimal trajectories and control functions for $5$ cities.} \\
    Optimal trajectories and controls for a problem involving $5$ cities. 
    The settings mirror those in \autoref{fig:combined_epidemic_behaviour_vaccination}, with the following specifications: $c_v = 0.01$, $c_h = 1000$, and $r_h = 0.1$. 
    The model allows for a weekly vaccination of up to $1/20$ of the susceptible population. 
    The maximum rate of vaccinated individuals is set at $v_1^{\max} = \dots = v_5^{\max} = 0.8/42$.}
\end{figure}

For a simulation with $K=5$ cities, refer to \autoref{fig:numerical_simulations_control_fig4}.
The results contained in the image are parallel to those in the two-city scenario in which the vaccines are preferentially allocated to the capital in contrast to other cities, vaccinating as fast as possible saturating the constraints.
During the fourth week, when the susceptible population is reduced to less than $60\%$ in the capital, the vaccination expands to the second city, which has the second highest transmission rate.
The behavior of the control function aligns again with the {\em bang-bang} type of solution.
In total, the experiment required approximately $3$ minutes and $31$ iterations.

\begin{figure}[!htbp]
    \centering
    \includegraphics[width=0.8\textwidth]{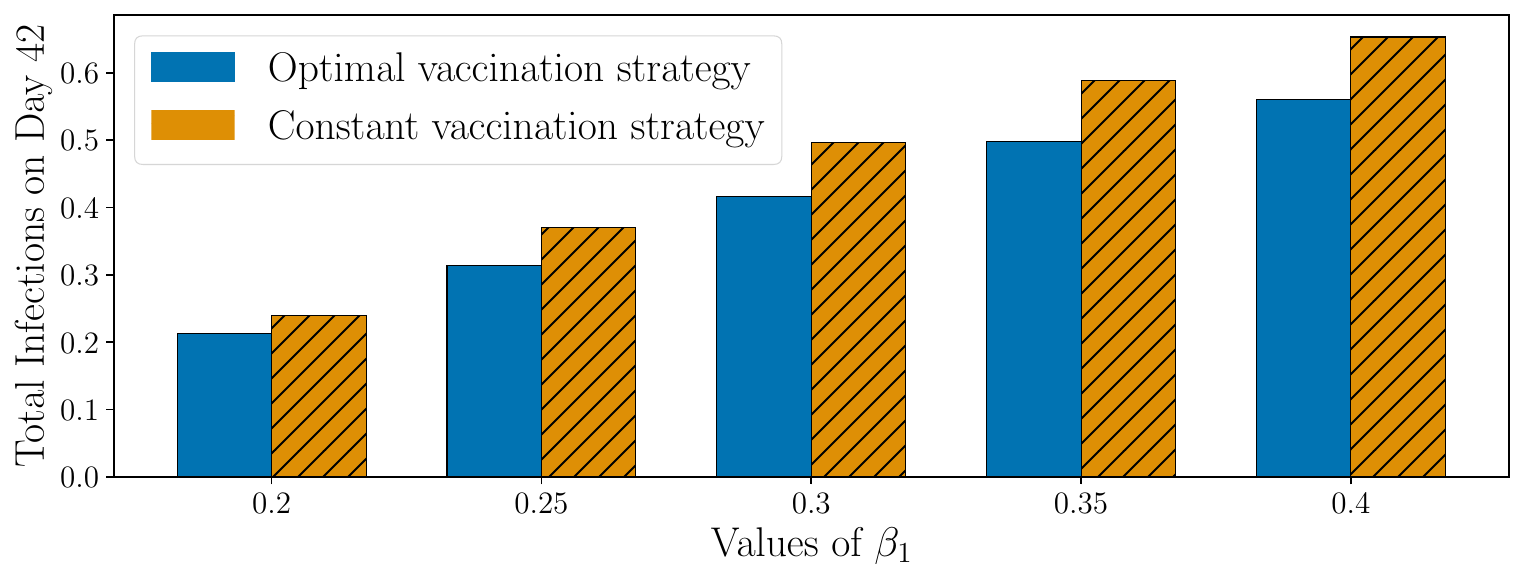}
    \caption{\label{fig:vaccination_strategy_comparison}\textbf{Comparison between optimal and constant vaccination strategies for different values of $\beta_1$}. \\
    Comparison among the total number of infections in the metropolitan area, which is calculated by integrating the new infections and taking the mean over the cities, for two scenarios.
    The first, in blue, solves the optimal control problem~\eqref{eq:optimal_control_problem}, and the second, in orange, chooses the best constant vaccination rate across all cities.
    }
\end{figure}

In addition to the analysis of the optimal control problem in multi-city interaction, we introduce a comparative visualization depicted in \autoref{fig:vaccination_strategy_comparison}. 
This figure presents a dual-bar plot comparison between the impact of two distinct vaccination strategies on the total number of infections across varying transmission rates of the capital city.
The transmission rate of the capital was chosen as a comparison factor since it drives most of the epidemic.
The bars in blue correspond to the application of the optimal control strategy as the solution of problem~\eqref{eq:optimal_control_problem}.
In orange, we consider a constant vaccination approach, where the vaccination rate is uniform in different cities and over time.
This comparison highlights that following the optimal strategy leads to fewer people having the disease during the epidemic. 
Additionally, the higher the transmission rate in the capital, the greater the impact of the optimal solution.

\subsection{Performance of feedback practical solution}

\begin{figure}[!hbtp]
    \centering
    \includegraphics[width=\textwidth]{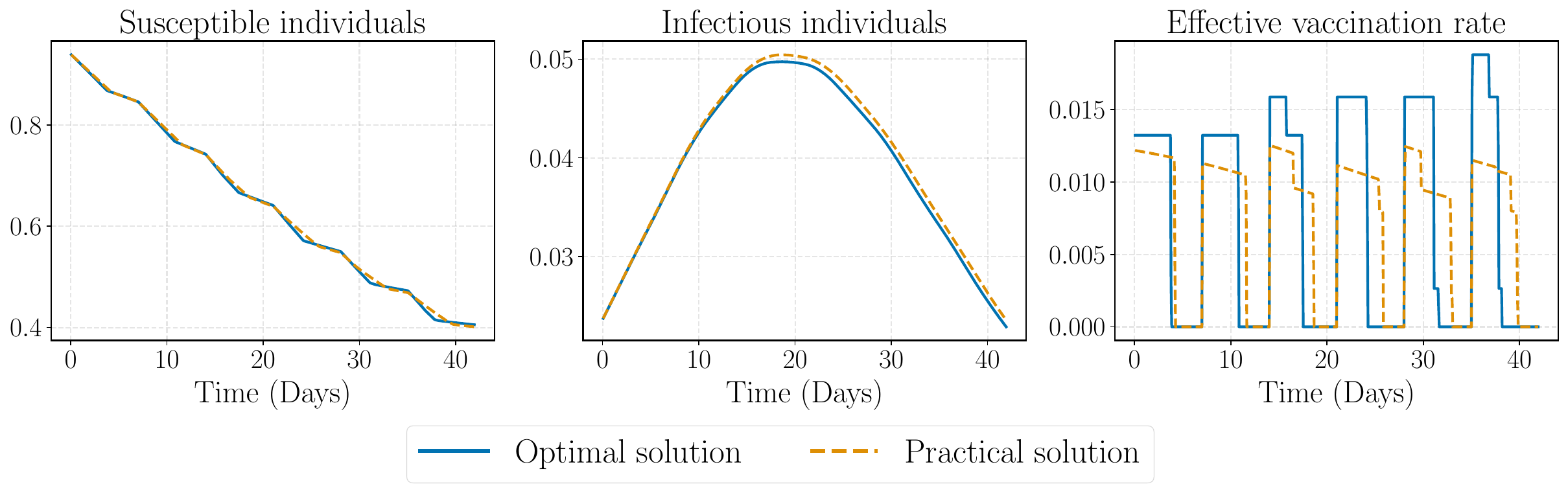}
    \caption{\label{fig:numerical_simulations_control_fig6}\textbf{Comparison between optimal and practical solution in the metropolitan area.} \\
    Optimal trajectories and effective vaccination rate for the metropolitan area considering the settings of \autoref{fig:numerical_simulations_control_fig4}.
    It compares the optimal solution with a more practicable solution, which is only based on the knowledge of vaccinated individuals.
    We observe that the effective vaccination rate here is the daily proportion of vaccinated individuals in the metropolitan area as a whole.}
\end{figure}

We bring to attention the fact that the optimal solution is hard to implement in real-world applications since the exact number of susceptible, infected, and recovered individuals is difficult to estimate.
Because of this, we propose a numerical simulation that substitutes the constraint $u_i(t)S_i(t)  \le \Di$ by
\[
u_i(t)(1 - V_i(t))  \le \Di
\]
which also leads to a bang-bang type of solution where we get $u_i^*(t)(1 - V_i^*(t)) \in \{0, \Di\}$ for $t\in [0, T]$.
If $\mu = 0$ (which is a reasonable approximation for short horizons), this solution is an admissible control to the original problem since
\[
\frac{d}{dt}(V_i + S_i + I_i) = \mu R_i - \gamma I_i \le 0,
\]
which implies
\[
V_i(t) + S_i(t) \le  V_i(t) + S_i(t) + I_i(t) \le V_i(0) + S_i(0) + I_i(0) \le 1
\]
and, therefore,
\[
u_i(t)(1 - V_i(t)) \ge u_i(t)S_i(t).
\]
Moreover, it is possible to apply in a real-world scenario, because $V_i(t)$ is naturally known by the Health System.
\autoref{fig:numerical_simulations_control_fig6} shows that this solution yields very similar results regarding the optimal trajectories since the curves of susceptible and infectious individuals end very closely.
The difference appears in the optimal vaccination policy since fewer vaccines can be administrated daily, but these are compensated by vaccinating during more days in the week.

\subsection{Evaluating Vaccination Strategies in the Rio de Janeiro Metropolitan Area}

\begin{figure}[!htbp]
\centering
\includegraphics[width=\textwidth]{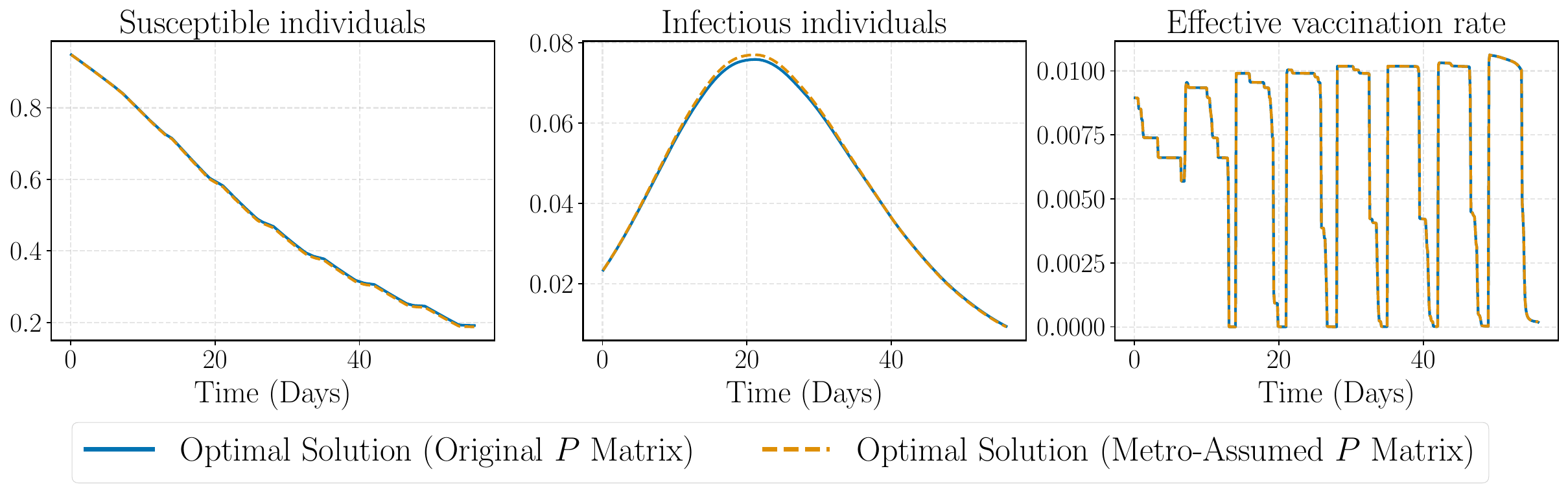}
\caption{\label{fig:numerical_simulations_control_rio_janeiro_comparison}\textbf{Comparative Analysis of Optimal Vaccination Strategies in the Rio de Janeiro Metropolitan Area.} \\
Optimal vaccination strategies for Rio de Janeiro, through both the original and approximated $P$ matrices. 
The settings are $\alpha = 0.64$, a maximum vaccination rate of $0.6/56$ per city, and weekly vaccine shipments for vaccinating up to $1/20$ of the total susceptible population.
The transmission rate $\beta$ is randomly chosen between $0$ and $0.3$, and sorted by city population size. 
Vaccination starts when the susceptible population falls below 95\% following the initial infection in the capital. 
The approximation of $P$ in a metropolitan context is calculated by normalizing $p_{i1}$ and $p_{ii}$ by its sum.}
\end{figure}

The last section of the numerical experiments considers real data on commuting patterns and population dynamics within the Rio de Janeiro metropolitan area~\cite{sebrae2013}. 
We examine the mobility matrix $P$, as depicted in \autoref{fig:rio_de_janeiro_pmatrix}. 
Despite the similarity of this matrix and the metropolitan structure outlined in \autoref{assumption:metropolitan_area}, the differences warrant further analysis.
\autoref{fig:numerical_simulations_control_rio_janeiro_comparison} shows a comparative study of the optimal vaccination strategies derived from both the actual matrix $P$ and its approximation considering the metropolitan hypothesis of \autoref{assumption:metropolitan_area}. 
This comparison reveals that the metropolitan structure approximation of the matrix $P$ yields an optimal solution that is closely similar to the one obtained using the original matrix.

\begin{remark}
    The parameter $\beta$ is proportional to the average contact rate of a random person.
    It is generally an increasing function of the population density. 
    Given the heterogeneous density in Rio de Janeiro’s metropolitan area, which includes large green spaces like the Tijuca Forest in the capital city, it would not be enough only to consider $\beta$ as proportional to the average density of each city. 
    For this reason, we consider random values for $\beta$ and sort them by population size.
\end{remark}

\begin{figure}[!hbtp]
\centering
\includegraphics[width=\textwidth]{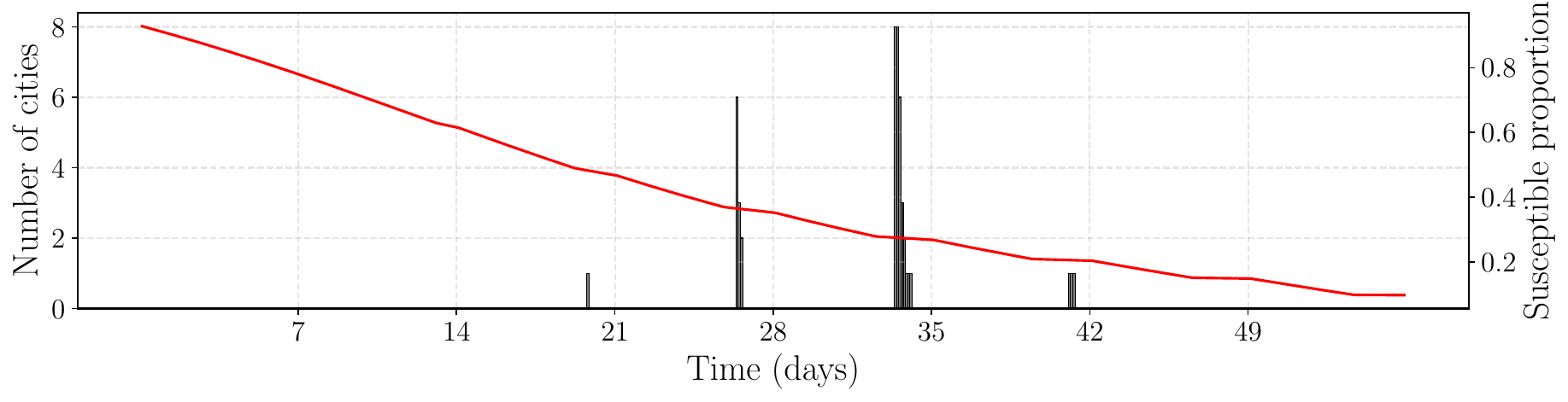}
\caption{\label{fig:numerical_simulation_rio_de_janeiro_vaccination}\textbf{Analysis of Effective Vaccination Rates in the Rio de Janeiro Metropolitan Area.} \\
The red graph represents the proportion of susceptible individuals in Rio de Janeiro city, while the gray bars show the number of cities vaccinating when the capital is not. 
The settings of this experiment are identical to those in \autoref{fig:numerical_simulations_control_rio_janeiro_comparison}.}
\end{figure}

Vaccination in the capital and the other cities is compared in \autoref{fig:numerical_simulation_rio_de_janeiro_vaccination}. 
This analysis tracks on a daily basis the number of cities that vaccinate while the capital is not, i.e., $u_1(t) = 0$ and some $u_i(t) \neq 0$ for $i \ge 2$.
We notice that most of the time, either the capital is vaccinating or none of the cities are.
The first deviation from this fact happens when the susceptible population is less than $60\%$, which corroborates our findings on the behavior of the optimal control function.

\begin{figure}[!htbp]
\centering
\includegraphics[width=\textwidth]{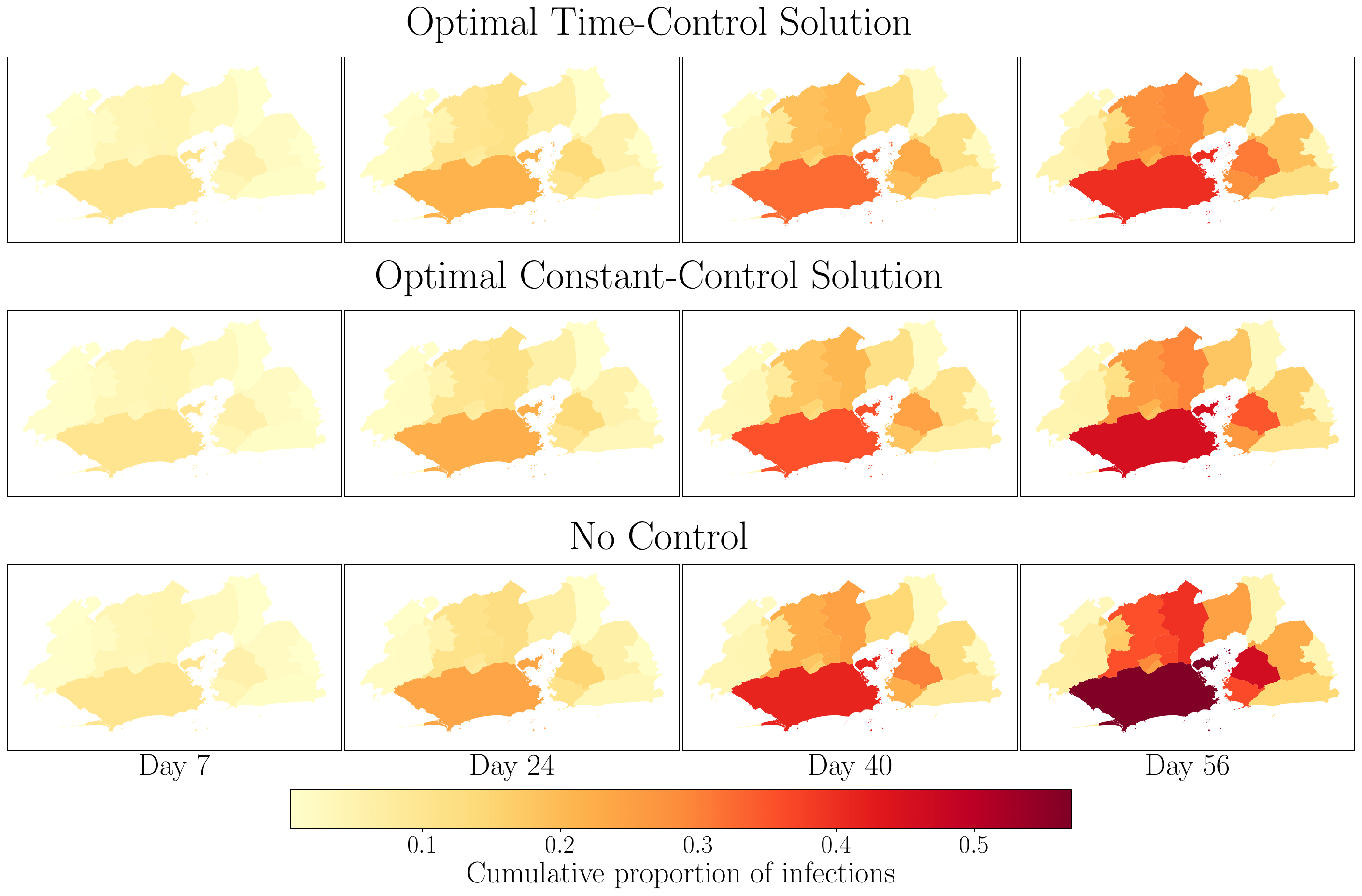}
\caption{\label{fig:numerical_simulation_rio_de_janeiro_geo}\textbf{Disease Progression in the Rio de Janeiro Metropolitan Area under Various Vaccination Scenarios.} \\
Cumulative proportion of infections via a heatmap, with the framework established in \autoref{fig:numerical_simulations_control_rio_janeiro_comparison}. 
Data sourced from \href{https://www.ibge.gov.br/geociencias/organizacao-do-territorio/malhas-territoriais/15774-malhas.html}{Brazilian Institute of Geography and Statistics (IBGE)}.}
\end{figure}

We conclude this section by presenting the evolution of the disease across the Rio de Janeiro metropolitan area in \autoref{fig:numerical_simulation_rio_de_janeiro_geo}. 
This figure compares three scenarios: the optimal solution of problem \eqref{eq:optimal_control_problem}, a uniform vaccination strategy across all cities and over time, and a scenario with no vaccination. 
We verify that the optimal solution induces a lower incidence of infections over the period. 
Moreover, it brings attention to the fact that smaller cities, such as Maricá, exhibit a higher proportion of infections under the optimal solution ($23\%$ over $56$ days) compared to the uniform vaccination strategy ($18\%$), due to the weighting of population sizes in the cost function \eqref{eq:cost_function}.

\section{Discussion}

Our research has led to significant advances in comprehending how epidemics behave within a metropolitan area, especially in relation to the most effective vaccination strategies. 
The key achievement of our study is the theoretical analysis of a mathematical model that combines commuting patterns with constrained optimal control to mimic and manage the propagation of infectious diseases. 
The study here presented aims to offer a realistic representation of disease transmission within densely populated urban regions.

As mentioned in the text, the application of optimal control in epidemiological models allows us to determine the most effective strategies for mitigating the impact of disease outbreaks while satisfying resource restrictions.
By incorporating commuting patterns into the model, we could capture the complex dynamics of disease spread in metropolitan areas. 
Ignoring them may lead to inaccurate predictions and ineffective control strategies.
Moreover, a strategy that does not consider several cities in its plan is doomed to failure due to mobility.

The literature review highlights the importance of integrating spatial heterogeneity and human/social behavior into epidemiological models. 
It also evidences the need for more realistic and detailed models tailored to specific research questions.

One of the key findings of this work regards the upper and lower bounds for the basic reproduction number, $\rzero$, in particular for the case of a metropolitan region.
However, a closed-form expression for it is still an open question since calculating the spectral radius of a diagonal plus one-rank matrix remains an open question in linear algebra.
We showed that $\rzero$ is close to $\rzero^1$ and most influenced by the infection rate in the capital city, $\beta_1$.
Thus, the upper bound may serve as an estimate for $\rzero$.
Efficient numerical methods for calculating $\rzero$ were discussed, but they are also open research in the field of perturbation theory.

The optimal control model possesses mixed control-state and pure-state constraints, in addition to control-affine dynamics and cost functional. 
This complex structure leads to many theoretical and numerical challenges, such as the occurrence of {\em singular arcs}, the presence of a multiplier that belongs to the space of measures, and others. 
We are developing the theoretical aspects related to the optimal control model \eqref{eq:optimal_control_problem} in another article (some results are already present in \cite{Moschen2023})

Our numerical simulations have provided valuable insights into the dynamics of the disease and the effectiveness of control measures. 
In particular, they showed that higher vaccination rates in the capital can significantly reduce the number of infections and the overall impact of the epidemic in the whole metropolitan area.
This underscores the importance of targeted vaccination strategies in controlling disease outbreaks.
We also highlighted how the infection rate of the capital drives most of the dynamics.

In conclusion, this study has contributed to our understanding of epidemic dynamics in metropolitan areas and the role of optimal control in mitigating disease outbreaks. 
It has also opened up several interesting directions for future research.
As we continue to grapple with the challenges posed by infectious diseases, studies like this one will be crucial in guiding our response and ensuring the health and well-being of our communities.

\section{Acknowledgments}

This research was mainly developed while the first author was a graduate student at the School of Applied Mathematics (FGV EMAp).
The authors kindly thank Guilherme Tegoni Goedert from FGV EMAp for providing the virtual machine used in this work; and to Luis Briceño Arias (Universidad Técnica Federico Santa María, Chile), Luis Almeida (Laboratoire Jacques-Louis Lions, France), and Maria do Rosário de Pinho (Universidade do Porto, Portugal) for their insightful discussions and valuable suggestions.
Both authors acknowledge the financial support from the School of Applied Mathematics (FGV EMAp), and Fundação Carlos Chagas Filho de Amparo à Pesquisa do Estado do Rio de Janeiro (FAPERJ) for the funding through process E-26/203.223/2017.
The second author further thanks the financial support of CNPq (Brazil) through process 310452/2019-8.

\bibliographystyle{abbrvnat}
\bibliography{biblio} 

\appendix

\makeatletter 
\renewcommand\thefigure{\thesection.\arabic{figure}}
\renewcommand\thetable{\thesection.\arabic{table}}
\makeatother

\section{The basic reproduction number for general compartmental models}\label{sec:r0_general_models}

We present the calculus of $\rzero$ within the framework of a general compartmental model, following the seminal work of \citeauthor{van2002reproduction} (\citeyear{van2002reproduction})~\cite{van2002reproduction}. 
They focus on autonomous systems, namely, those whose right-hand side does not explicitly depend on $t$.
Let $x \in \R_{\geq 0}^{m+n}$ (real vectors with non-negative entries) be the number (or proportion) of individuals in $m+n$ compartments, where the first $m$ are infected states and the remaining $n$ are non-infected states.
The rate of new infections in the $i$th infected compartment is denoted by $\mathcal{F}_i$, the rate of transfer into the $i$th compartment (except new infections) by $\mathcal{V}_i^+$, and the rate of transfer out of the $i$th compartment by $\mathcal{V}_i^-$.
The net transfer is then given by $\mathcal{V}_i = \mathcal{V}_i^{-} - \mathcal{V}_i^{+}$. 
The disease transmission model consists of the equations
\[
\begin{split}
	x'_i &= f_i(x) = \mathcal{F}_i(x) - \mathcal{V}_i(x), \text{ for } i=1,\dots,m+n. \\
\end{split}
\]
The choice of the infected and non-infected compartments depends on the model's interpretation.

A {\em disease-free state\/} is a state $x \in \R_{\geq 0}^{m+n}$ such that $x_i = 0$ for $i=1,\dots,m$, and a {\em disease-free equilibrium\/} (DFE) is a disease-free state that is an asymptotically stable equilibrium.
We ensure the model's well-posedness and the existence of such equilibrium, we make the following assumptions:
\begin{enumerate}[label = (A\arabic*)]
    \item The transfer of individuals between compartments is non-negative: if $x \in \R_{\geq 0}^{m+n}$, then \[\mathcal{F}_i(x), \mathcal{V}_i^+(x), \mathcal{V}_i^{-}(x) \ge 0.\]
    \item There is no transfer out of an empty compartment: if $x = 0$, then $\mathcal{V}_i^-(x) = 0$.
    \item Non-infected compartments do not receive new infections: $\mathcal{F}_i(x) = 0$ for $i > m$.
    \item The set of the disease-free states is invariant: if $x_i = 0$ for $i=1,\dots,m$, then $\mathcal{F}_i(x) = \mathcal{V}_i^+(x) = 0$ for $i=1,\dots,m$.
    \item The DFE is stable in the absence of new infections: if $\mathcal{F}(x) = 0$ and $x_0$ is a DFE, $Df(x_0)$ is a Hurwitz matrix. 
\end{enumerate}

From these assumptions and if $x_0$ is a DFE, we obtain that 
\[
D \mathcal{F}(x_0) = \begin{bmatrix}
    F & 0 \\ 0 & 0
\end{bmatrix}, \quad D \mathcal{V}(x_0) = \begin{bmatrix}
    V & 0 \\ J_3 & J_4
\end{bmatrix}, 
\]
where $F$ is non-negative, $V$ is a non-singular $M$-matrix and the eigenvalues of $J_4$ have positive real part~\cite[Lemma 1]{van2002reproduction}.
A square matrix $B$ is an {\em $M$-matrix}  
if it can be expressed as $B = sI - P$, where $P$ is a matrix with non-negative elements, and $s$ is a real scalar such that $s \geq \rho(P)$.  
If $s > \rho(P)$, then $B$ is a {\em non-singular $M$-matrix}.
If $s = \rho(P)$, it is a {\em singular $M$-matrix}.
The matrix $FV^{-1}$ is named the {\em next generation matrix\/} for the model and 
\begin{equation}\label{eq:rzero_definition}
    \rzero = \rho(FV^{-1}),
\end{equation} 
in which $\rho(A)$ is the spectral radius of the matrix $A$.
The $(i,j)$ entry of matrix $FV^{-1}$ estimates how many new infections are expected to arise in compartment $i$ if an infected individual is introduced into compartment $j$.
This definition implies that $x_0$ is asymptotically stable if $\rzero < 1$, but unstable if $\rzero > 1$~\cite[Theorem 2]{van2002reproduction}.

\subsection{Computing the basic reproduction number for our model}\label{sec:appendix-r0}

Following the notation of~\autoref{sec:r0_general_models}, we reorder the compartments as follows:
\[
x = (I_1, \dots, I_K, S_1, \dots, S_K, R_1, \dots, R_K).
\] 
The rate of new infections is 
\[
\mathcal{F}_i = \alpha \beta_i S_i I_i + (1-\alpha) S_i \sum_{j=1}^K \beta_j p_{ij} I_j^{\mathrm{eff}}, \text{ for } i = 1,\dots, K,
\]
and zero for non-infected compartments. 
The rate of transfer between compartments is defined as
\[
\mathcal{V}_k = \begin{cases}
    (\gamma + \mu) I_i, &1 \le k \le K, \, i=k,\\
    \alpha \beta_i S_i I_i + (1-\alpha) S_i \sum_{j=1}^K \beta_j p_{ij} I_{j}^{\mathrm{eff}} + \mu S_i - \mu, &K+1 \le k \le 2K, \, i=k-K, \\
    \mu R_i - \gamma I_i, &2K+1 \le k \le 3K, \, i=k-2K.
\end{cases}    
\]
Utilizing the auxiliary result
\[
\frac{d I_k^{\mathrm{eff}}}{d I_j} = \frac{p_{jk} n_j}{P_k^{\mathrm{eff}}},
\]
we proceed to calculate matrices $F$ and $V$. The elements of these matrices are given by
\[
F_{ij} = \alpha \beta_i S_i I_i \delta_{ij} + (1-\alpha) S_i \sum_{k=1}^K \beta_k p_{ik} p_{jk} \frac{n_j}{P_k^{\mathrm{eff}}}, 
\]
\[
V_{ij} = \gamma + \mu,
\]
where $\delta_{ij} = 1$ if $i=j$ and $0$ otherwise. We can represent the matrix $F$ as 
\begin{equation}\label{eq:matrix_A}
    F = \alpha \mathcal{S} B + (1-\alpha) \mathcal{S} P B E^{-1} P^T N = \mathcal{S}\left[\alpha B + (1-\alpha) P B E^{-1} P^T N\right],
\end{equation}
where $\mathcal{S}, N, B$ and $E$ are diagonal matrices, such that ${\mathcal{S}}_{ii} = S_i$, $N_{ii} = n_i$, $B_{ii} = \beta_i$ and $E_{ii} = P_i^{\mathrm{eff}}$.
Intuitively, the entry $(i,j)$ of $F$ is the rate at which infected individuals in city $j$ contribute to new infections in city $i$.
Therefore, the sum $\sum_{j=1}^K A_{ij}$ quantifies the total rate of new infections in city $i$.
Then 
\[
FV^{-1} = \frac{1}{\gamma + \mu} F,
\]
the eigenvalues of $FV^{-1}$ are the eigenvalues of $F$ divided by $\gamma + \mu$, and $\rzero = \rho(F)/(\gamma + \mu)$.
We can readily see that matrix $F$ is equivalent to a symmetric matrix, which implies the following lemma.

\begin{lemma}\label{lemma:real_eigenvalues}
    The eigenvalues of $FV^{-1}$ are real.
\end{lemma}

As far as we know, there is no closed-form expression for the spectral radius of $F$ as a function of the parameters of the model. 
However, we can simplify equation~\eqref{eq:matrix_A} by applying~\autoref{assumption:metropolitan_area} which defines a specific format for $P$.
Consequently, we calculate that
\[
PBE^{-1} = \begin{bmatrix}
    \frac{\beta_1}{P_1^{\mathrm{eff}}} & 0 & 0 & \cdots & 0 \\
    \frac{\beta_1}{P_1^{\mathrm{eff}}} p_{21} & \frac{\beta_2}{P_2^{\mathrm{eff}}} p_{22} & 0 & \cdots & 0 \\
    \frac{\beta_1}{P_1^{\mathrm{eff}}} p_{31} & 0 & \frac{\beta_3}{P_3^{\mathrm{eff}}} p_{33} & \cdots & 0 \\
    \vdots & \vdots & \vdots & \ddots & \vdots \\
    \frac{\beta_1}{P_1^{\mathrm{eff}}} p_{K1} & 0 & 0 & \cdots & \frac{\beta_K}{P_K^{\mathrm{eff}}} p_{KK}
\end{bmatrix}
\]
and, subsequently,
\[
PBE^{-1}P^T N = \frac{\beta_1}{P_1^{\mathrm{eff}}} \bar{p} \bar{p}^T N + \Lambda,     
\]
where $\bar{p} = (1, p_{21}, \dots, p_{K1})$ and $\Lambda = \operatorname{diag}{\left(\beta_k p_{kk} (1 - \delta_{k1})\right)}_{k=1,\dots,K}$, with $P_k^{\mathrm{eff}} = p_{kk} n_k$ for $k > 1$.
Therefore, we can express the matrix $F$ from~\eqref{eq:matrix_A} as 
\begin{equation}\label{eq:expression2_F}
    F = \mathcal{S}\left[\alpha B + (1-\alpha) \left( \frac{\beta_1}{P_1^{\mathrm{eff}}} \bar{p} \bar{p}^T N + \Lambda\right) \right]
    = \mathcal{S}[\alpha B + (1-\alpha)\Lambda] + (1-\alpha)\frac{\beta_1}{P_1^{\mathrm{eff}}}(\mathcal{S}\bar{p}) (\bar{p}^T N),
\end{equation}
which is a sum of a diagonal matrix and a product of vectors.
The problem of finding the eigenvalues of matrices with this structure, referred to as {\em diagonal plus rank-one matrix}, such as matrix $F$, is well-known in the literature, and there are specific algorithms to solve it, such as~\cite{golub1973some, stor2015forward}.
An inequality for $\rzero$ can be computed through Weyl's inequality by considering this structure.

The unique disease-free equilibrium of system~\eqref{eq:robot-dance-model} sets $S_i = 1$ and $I_i = R_i = 0$ for all cities $i=1,\dots,K$, which simplifies $\mathcal{S}$ to the identity matrix.
In the modified model~\eqref{eq:robot-dance-model-vaccination} with a constant vaccination rate in the population, the dynamics of compartments $I_i$ are unchanged, which leads to the same calculation of the basic reproduction number.
However, the disease-free equilibrium is redefined as
\[
\mu - \mu S_i - u_i S_i = 0 \implies S_i = \frac{\mu}{\mu + u_i} \quad \text{ and } \quad R_i = \frac{u_i}{\mu + u_i},
\]
besides $I_i = 0$.
Therefore the only change is on the diagonal matrix $\mathcal{S}$.


\section{Proofs}\label{sec:appendix-proof}

This section includes all the proofs of results given throughout the text.

\begin{proposition}[Positive Invariance]\label{prop:positive-invariance}
    The set
    \[
    \mathcal{C} \coloneqq \{X \in \R_{\geq 0}^{K \times 3} : X_{i1} + X_{i2} + X_{i3} = 1, \text{ for } i=1,\dots,K\}    
    \]
    is positively invariant under the flow of system~\eqref{eq:robot-dance-model}.
\end{proposition}

\begin{proof}[Proof]\phantomsection\label{proof:positive-invariance}
    Firstly, by the smoothness of the system, existence and uniqueness of solution $[\boldsymbol{S}(t), \boldsymbol{I}(t), \boldsymbol{R}(t)]$ in $[0, T]$ such that $[\boldsymbol{S}(0), \boldsymbol{I}(0), \boldsymbol{R}(0)] \in \mathcal{C}$ is guaranteed.
    By the uniqueness, since $\boldsymbol{\tilde{I}}(t) = \boldsymbol{\tilde{R}}(t) = 0$ is also a solution over $[0, T]$, we have $I_i(t), R_i(t) \ge 0$ for all $i=1,\dots,K$.
    Moreover, notice that 
    \[
    S_i'(t) \ge -S_i(t)\left[\alpha \beta_i I_i + (1-\alpha)\sum_{j=1}^K \beta_j p_{ij} I_j^{\mathrm{eff}} + \mu\right],
    \]
    which implies, by Gronwall's Inequality, that
    \[
    -S_i(t) \le -S_i(0)\exp\left\{-\int_0^T \alpha \beta_i I_i(s) + (1-\alpha)\sum_{j=1}^K \beta_j p_{ij} I_j^{\mathrm{eff}}(s) \, ds - \mu\right\} \le 0.
    \]
    This proves the non-negativity of the solution.
    Finally it is straightforward that for each $i$, $S_i'(t) + I_i'(t) + R_i'(t) = 0$, which implies that $S_i(t) + I_i(t) + R_i(t) = S_i(0) + I_i(0) + R_i(0) = 1$ for all $t \in [0, T]$.
    If we consider the parameter $\alpha$ as a constant by parts function, notice that the result is proven by induction in each interval in which $\alpha(\cdot)$ is constant.
\end{proof}

\begin{proof}[Proof of \autoref{thm:bounds_rzero_generalP}]\phantomsection\label{proof:bounds_rzero_generalP}
    Formula~\eqref{eq:isolated_cities} is derived from the basic reproduction number from a simple SIR model.
    By considering the expression~\eqref{eq:matrix_A} for matrix $F$, for each $1 \le i,j \le K$,
    \[
    \begin{split}
        F_{ij} &= S_i \left[\alpha \beta_i \delta_{ij} + (1-\alpha){\left(PBE^{-1} P^T N\right)}_{ij}  \right] \\
        &= S_i\left[\alpha \beta_i \delta_{ij} + (1-\alpha) \sum_{k=1}^K {\left(P B E^{-1}\right)}_{ik} {\left(P^T N\right)}_{kj} \right] \\
        &= S_i\left[\alpha \beta_i \delta_{ij} + (1-\alpha) \sum_{k=1}^K p_{ik}\beta_k p_{jk} n_j / P_k^{\mathrm{eff}} \right],
    \end{split}
    \]
    where  $\delta_{ij} = 1 \iff i = j$ and $\delta_{ij} = 0$ otherwise.
    Assuming $\alpha > 0$ and that for all cities $S_i > 0$ and $\beta_i > 0$, we infer that $F_{ij} > 0$.
    This means that $F$ is a positive matrix and satisfies Perron-Frobenius's theorem.
    We conclude that there is $r > 0$ such that $r$ is an eigenvalue of $F$ and $\rho(F) = r$. 
    Furthermore, 
    \[
    \min_i \sum_{j=1}^K F_{ij} \le r \le \max_i \sum_{j=1}^K F_{ij}.
    \]
    Calculating,
    \[
    \begin{split}
        \sum_{j=1}^K F_{ij} &= S_i\left[\alpha \beta_i + (1-\alpha) \sum_{j=1}^K \sum_{k=1}^K p_{ik}\beta_k / P_k^{\mathrm{eff}} p_{jk} n_j \right]  \\
        &= S_i\left[\alpha \beta_i + (1-\alpha) \sum_{k=1}^K  p_{ik}\beta_k / P_k^{\mathrm{eff}} \sum_{j=1}^K p_{jk} n_j \right]  \\
        &= S_i\left[\alpha \beta_i + (1-\alpha) \sum_{k=1}^K  p_{ik}\beta_k \right],  \\
    \end{split}        
    \]
    which implies that 
    \begin{equation*}
        \min_i S_i\left[\alpha \beta_i + (1-\alpha) \sum_{k=1}^K  p_{ik}\beta_k \right] \le \rho(F) \le \max_i S_i\left[\alpha \beta_i + (1-\alpha) \sum_{k=1}^K  p_{ik}\beta_k \right]. 
    \end{equation*}
    At the DFE, where $S_i = 1$ for all cities $i$, we conclude that
    \[
    \frac{ \min_i \alpha \beta_i + (1-\alpha) \sum_{k=1}^K  p_{ik}\beta_k  }{\gamma + \mu}  \le \rzero \le \frac{ \max_i \alpha \beta_i + (1-\alpha) \sum_{k=1}^K  p_{ik}\beta_k}{\gamma + \mu}
    \]
    or, in terms of the basic reproduction numbers of the isolated cities,
    \[
    \min_i \alpha \rzero^i + (1-\alpha) \sum_{k=1}^K p_{ik} \rzero^k \le \rzero \le \max_i \alpha \rzero^i + (1-\alpha) \sum_{k=1}^K p_{ik} \rzero^k.   
    \]
\end{proof}

\begin{proof}[Proof of~\autoref{cor:lower_upper_bound}]\phantomsection\label{proof:lower_upper_bound}
    Immediate since $\alpha + (1-\alpha)\sum_{k=1}^K p_{ik} = 1$ and, therefore, $w_i$ is a convex combination of $\rzero^1, \dots, \rzero^K$.
\end{proof}

\begin{proof}[Proof of~\autoref{thm:bounds_rzero_metropolitan}]\phantomsection\label{proof:bounds_rzero_metropolitan}
    Following the expression~\eqref{eq:expression2_F}, rewrite the matrix $F$ as 
    \[
    F = N^{-1/2}\mathcal{S}^{1/2}[\alpha \mathcal{S} B + (1-\alpha) \mathcal{S} \Lambda + (1-\alpha)\frac{\beta_1}{P_1^{\mathrm{eff}}} \mathcal{S}^{1/2} N^{1/2} \bar{p} \bar{p}^T N^{1/2}\mathcal{S}^{1/2}]\mathcal{S}^{-1/2}  N^{1/2},
    \]
    which is similar, and therefore has the same eigenvalues, to
    \[
    \bar{F} = \alpha \mathcal{S} B + (1-\alpha) \mathcal{S} \Lambda + (1-\alpha)\frac{\beta_1}{P_1^{\mathrm{eff}}} \mathcal{S}^{1/2} N^{1/2} \bar{p} {\left(\mathcal{S}^{1/2} N^{1/2} \bar{p}\right)}^{T}.
    \]
    By setting $\mathcal{S}$ as the identity matrix (DFE) and utilizing the symmetry of the matrices, we can apply Weyl's inequality to obtain
    \[
    \rho(\alpha B + (1-\alpha) \Lambda) \le \rho(F) \le \rho(\alpha B + (1-\alpha) \Lambda) + (1-\alpha)\frac{\beta_1}{P_1^{\mathrm{eff}}} {(N^{1/2} \bar{p})}^T N^{1/2} \bar{p},
    \]
    since the only non-zero eigenvalue of a matrix $vv^T$ is $v^T v$.
    We finally set $\xi = \rho(\alpha B + (1-\alpha) \Lambda)$ and calculate ${(N^{1/2} \bar{p})}^T N^{1/2} \bar{p} = \sum_{i=1}^K n_i p_{i1}^2$.
\end{proof}

\begin{proposition}[Positive Invariance with control]\label{prop:positive-invariance-control}
    The region 
    \[
    \mathcal{C} = \{X \in \R_{\geq 0}^{K \times 4} : X_{i1} + X_{i2} + X_{i3} = 1, \text{ for } i=1,\dots,K\}    
    \]
    is positively invariant under the flow of system~\eqref{eq:robot-dance-model-vaccination}, for each measurable function $u$.
\end{proposition}

\begin{proof}[Proof]\phantomsection\label{proof:positive-invariance-control}
    Let $u$ be a measurable function in $[0, T]$ and suppose it assumes values on a compact set $U$.
    Since the dynamic is smooth, there is a unique solution $[\boldsymbol{S}(t), \boldsymbol{I}(t), \boldsymbol{R}(t), \boldsymbol{V}(t)]$ for $t \in [0, T]$~\cite{bressan2007introduction}. 
    As similarly proven for \autoref{prop:positive-invariance}, 
    \begin{equation}\label{eq:aux-inequality}
    -S_i(t) \le -S_i(0)\exp\left\{-\int_0^T \alpha \beta_i I_i(s) + (1-\alpha)\sum_{    j=1}^K \beta_j p_{ij} I_j^{\mathrm{eff}}(s) + u_i(s) \, ds - \mu\right\} \le 0,
    \end{equation}
    for each $i=1,\dots,K$.
    By uniqueness, $I_i(t) \ge 0$ for each $i=1,\dots, K$ since the solution cannot cross the solution with no infections.
    Moreover, $R_i'(t) \ge -\mu R_i(t)$, which results, by Gronwall's Inequality, in 
    \[
    -R_i(t) \le -R_i(0)e^{-\mu t} \le 0.
    \]
    Finally $V_i'(t) \ge 0$ for every $t \in [0,T]$, which implies $V_i(t) \ge V_i(0) = 0$ for every $t \in [0,T]$.
    We conclude by observing that 
    \[
    \frac{d(S_i+I_i+R_i)}{dt} = 0 \implies (S_i+I_i+R_i)(t) = (S_i+I_i+R_i)(0) = 1,
    \]
    for all $t > 0$ and $i=1,\dots,K$.
    Supposing that $U$ is compact was without loss of generality because we impose that $u_i(t) S_i(t) \le \Di$ for each $i$ and each $t$.
    Therefore, 
    \[
    \Di \ge u_i(t) S_i(t) \ge u_i(t) \min_{t \in [0,T]} S_i(t) \implies 0 \le u_i(t) \le \Di/\min_{t \in [0,T]} S_i(t)
    \]
    and $\min_{t \in [0,T]} S_i(t) > 0 \iff S_i(0) > 0$ by inequality~\eqref{eq:aux-inequality}.
\end{proof}

\begin{proof}[Proof of~\autoref{lemma:real_eigenvalues}]\phantomsection\label{proof:real_eigenvalues}
    The matrix $F$ can be rewritten as 
    \[
    F = N^{-1/2} \mathcal{S}^{1/2}[\alpha B \mathcal{S} + (1-\alpha) \mathcal{S}^{1/2} N^{1/2}  P B E^{-1} P^T N^{1/2} \mathcal{S}^{1/2}] \mathcal{S}^{-1/2} N^{1/2},
    \]
    which is similar, in the sense of matrices, to $\alpha B \mathcal{S} + (1-\alpha) \mathcal{S}^{1/2} N^{1/2}  P B E^{-1} P^T N^{1/2} \mathcal{S}^{1/2}$.
    The matrix $B \mathcal{S}$ is diagonal, while $\mathcal{S}^{1/2} N^{1/2}  P B E^{-1} P^T N^{1/2} \mathcal{S}^{1/2}$ is symmetric.
    Therefore $F$ is similar to a symmetric matrix and, consequently, it is symmetric.
    Consequently, the eigenvalues of $F$ are real and so are the eigenvalues of $FV^{-1}$.
\end{proof}

\section{Additional simulations and figures}
\label{appendix:figures}

In this section, we present additional results that are less central to the overall work but may be of interest to the audience.

\subsection{On the effect of some parameters in \texorpdfstring{$\rzero$}{rzero}}

\begin{figure}[!htbp]
    \centering
    \includegraphics[width=\textwidth]{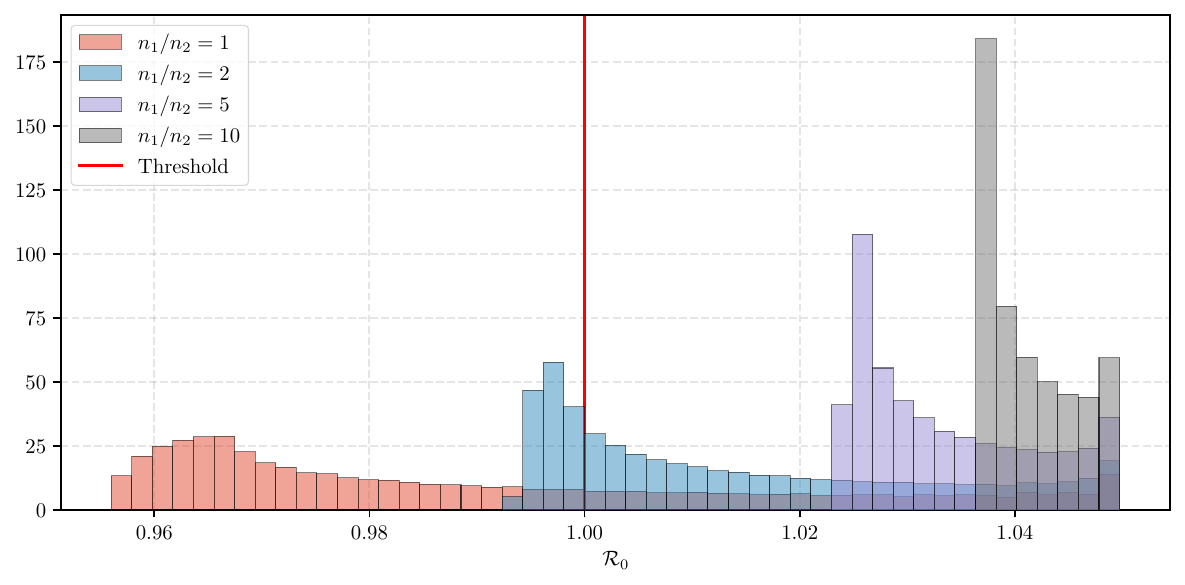}
    \caption{\label{fig:r0_function_alpha_p21_smaller_beta}\textbf{Histogram of the possible values of $\mathcal{R}_0$ as function of $\alpha$ and $p_{21}$ for two cities}:
    histogram of the values of $\rzero$ generated by randomly selecting $\alpha \in (0,1)$ and $p_{21} \in (0,0.9)$. 
    We consider four scenarios for the ratio $n_1/n_2$, $(1,2,5,10)$. 
    The other parameters are $\beta_1 = 0.15$, $\gamma = 1/7$ and $\mu = 3.6\cdot 10^{-5}$, the same as in \autoref{fig:r0_function_alpha_p21}.
    }
\end{figure}

Throughout the text, we have observed that the transmission rate primarily influences the spread of the disease in the population. 
Considering a scenario involving a metropolitan area with two cities, \autoref{fig:r0_function_alpha_p21_smaller_beta} illustrates the impact of parameters $\alpha$ and $p_{21}$ on $\rzero$. 
The histogram displays possible values of $\rzero$ as a function of random combinations of these parameters.
Notably, it reveals that higher values of the ratio $n_1/n_2$ shrink the distribution, as a byproduct of the fact that as $n_1/n_2$ approaches infinity, $\rzero$ converges to $\rzero^1$. 
By setting $\beta_1$ close to $\gamma$, then $\rzero \approx 1$, we observe that for smaller values of the ratio $n_1/n_2$, the mobility parameters may affect the threshold $\rzero > 1$, changing the behavior of the disease.

\subsection{Epidemic values for different structures of the transition matrix}

In the context of multiple cities, we consider a scenario of $K=5$ cities.
We first examine five distinct mobility scenarios: (I) metropolitan area where $10\%$ of the residents of each city work in the capital; (II) metropolitan area where $40\%$ of each city works in the capital; (III) all cities are interconnected, where 60\% of the residents work in their city, while the remaining are distributed evenly across the other four cities; (IV) a variant of a metropolitan area where 30\% of each city's residents work in the capital and 10\% of the capital's workforce is employed in the other cities; (V) no commuting between cities.
Our findings are summarized in \autoref{table:summary_scenarios_matrix_P}.

\begin{table}[!htbp]
\centering
\begin{tabular}{l c c c c c c c }
\multirow{2}{*}{Metrics} & \multirow{2}{*}{Scenarios} & \multicolumn{5}{c}{Cities} & \multirow{2}{*}{Aggregated} \\ \cline{3-7}
 &  & 1 & 2 & 3 & 4 & 5 &  \\ \toprule{}
\multirow{4}{*}{Peak size (\%)} & (I) & 27.0 & 14.1 & 9.3 & 5.7 & 3.7 & 20.5 \\
 & (II) & 26.3 & 18.4 & 15.3 & 12.5 & 10.2 & 22.5 \\
 & (III) & 24.2 & 17.3 & 14.7 & 12.5 & 10.2 & 21.0 \\ 
 & (IV) & 25.5 & 17.7 & 14.3 & 12.7 & 9.8 & 21.7 \\
 & (V) & 27.5 & 10.9 & 4.5 & 0.0 & 0.0 & 19.2 \\[1ex]
\multirow{4}{*}{Peak day} & (I) & 39 & 53 & 55 & 52 & 49 & 40 \\ 
 & (II) & 40 & 46 & 47 & 47 & 47 & 41 \\ 
 & (III) & 44 & 49 & 50 & 50 & 50 & 45 \\ 
 & (IV) & 41 & 47 & 49 & 50 & 50 & 45 \\ 
 & (V) & 38 & 81 & 134 & --- & --- & 38 \\[1ex]
\multirow{4}{*}{Duration (days)} & (I) & 146 & 188 & 218 & 234 & 191 & 191 \\
 & (II) & 145 & 164 & 170 & 172 & 166 & 157 \\
 & (III) & 151 & 168 & 172 & 171 & 168 & 161 \\
 & (IV) & 147 & 167 & 175 & 167 & 164 & 160 \\
 & (V) & 137 & 228 & 350 & --- & --- & 306 \\[1ex]
\multirow{4}{*}{Attack rate (\%)} & (I) & 91.4 & 72.9 & 58.1 & 37.4 & 20.4 & 82.5 \\
 & (II) & 91.2 & 77.6 & 69.0 & 58.3 & 47.0 & 85.2 \\
 & (III) & 89.8 & 75.5 & 67.4 & 58.3 & 47.9 & 83.7 \\
 & (IV) & 90.7 & 76.6 & 66.9 & 57.5 & 44.9 & 84.3 \\
 & (V) & 91.5 & 70.6 & 50.8 & 0.0 & 0.0 & 80.4 \\ \bottomrule{}
\end{tabular}

\caption{\label{table:summary_scenarios_matrix_P}\textbf{Simulation results for different structures of the transition matrix}:
presents the comparison of Peak Size (maximum proportion of infectious individuals), Peak Day (day of the peak size), Duration (time from epidemic onset to the day when the proportion of infectious individuals achieves $10^{-5}$), and Attack Rate (proportion of individuals who contract the disease during $350$ days of the epidemic) across different mobility scenarios.
The aggregated column represents the metrics considering all cities together.
For this experiment we $\beta = (0.4, 0.25, 0.2, 0.15, 0.1)$, $\alpha = 0.64$ and population sizes $10^5\cdot(50, 10, 10, 1, 1)$.
Lastly, the initial conditions are $I_1(0) = I_2(0) = I_3(0) = 10^{-4}$ and no recovered individuals.
}
\end{table}

\subsection{Evaluating the sharpness of the bound for \texorpdfstring{$\rzero$}{rzero}}

Considering a scenario with two cities, we perform a simulation over $N=100,000$ iterations which randomly assigns values for $\beta_2 \in (0.01, 0.2)$, $\beta_1 \in (\beta_2, 0.6)$, $\alpha \in (0, 1)$, $p_{21} \in (0, 0.9)$, and $n_1 \in (1,10) \cap \mathbb{Z}$. 
The simulation then computes the upper and lower bounds using both general (\autoref{thm:bounds_rzero_generalP}) and metropolitan (\autoref{thm:bounds_rzero_metropolitan}) bounds, recording whether the latter provides a tighter bound than the former. 
The results reveal that the metropolitan method offers better bounds in 50.0\% of the cases. 
In 18.1\% of the cases, the metropolitan method provides a better upper bound but not a better lower bound, while in 31.9\% of the cases, the general method provides better bounds. 
In the remaining 18.0\% of cases, the bounds are equal.
When we assume $\alpha \ge 0.5$, the bounds from \autoref{thm:bounds_rzero_metropolitan} are better in 87.1\% of the simulations and better only for the upper bounds in 4.9\% of the simulations.
The general method provides better bounds in only 8.1\% of the simulations.

In the five-city scenario, we perform a simulation that runs for $N=500,000$ iterations, with each iteration randomly generating values for $\beta \in (0.01, 0.6)$, $\alpha \in (0, 1)$, $p \in {(0, 0.9)}^4$ being the transition probabilities from each of the other four cities to the capital, and $n \in {(1,50)}^5 \cap \mathbb{Z}$ the population sizes. 
We sort the vector $n$ in descending order. 
The code then calculates the upper and lower bounds, the same as before. 
The results of the simulation indicate that the metropolitan method provides better bounds in 48.9\% of the cases. 
In 46\% of the cases, the metropolitan method provides a better lower bound but not a better upper bound, while in 3.8\% of the cases, the general method provides better bounds. 
When we assume $\alpha \ge 0.5$, the bounds of \autoref{thm:bounds_rzero_metropolitan} are better in 80.6\% of the simulations and better only for the lower bounds in 19.4\% of the simulations.
No recorded simulation had better bounds for the general method, arguing in favor of tighter bounds for $K=5$ cities.

\subsection{Impact of vaccination in the peak size}

\begin{figure}[!bth]
    \centering
    \begin{subfigure}[b]{0.9\textwidth}
        \includegraphics[width=\textwidth]{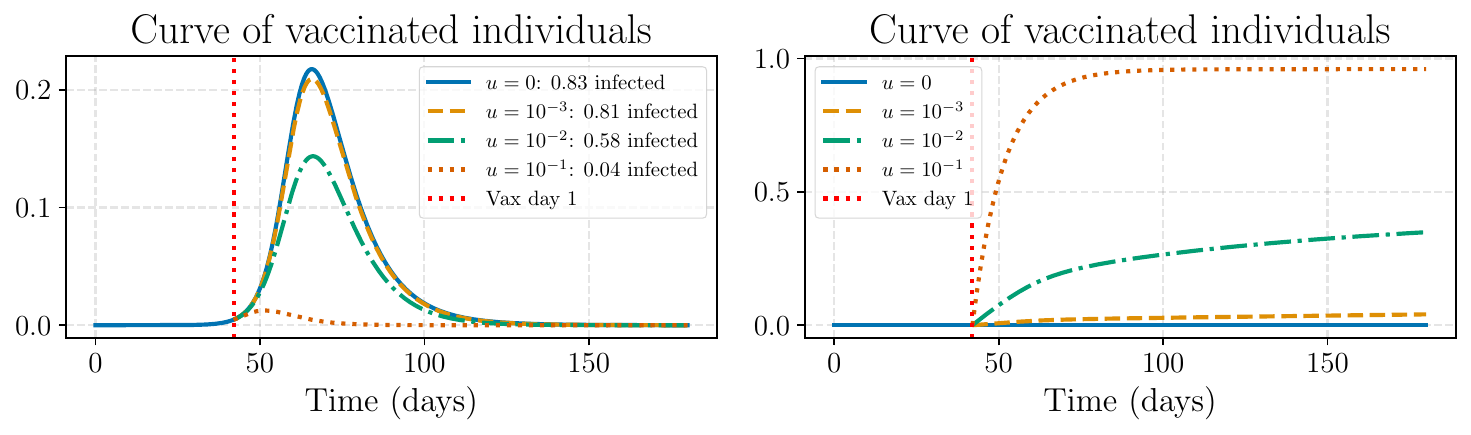}
        \caption{\label{fig:epidemic_behaviour_vaccination}Dynamics of infectious and vaccinated individuals in the metropolitan area}
    \end{subfigure}
    \vspace{20pt}
    \begin{subfigure}[b]{0.9\textwidth}
        \includegraphics[width=\textwidth]{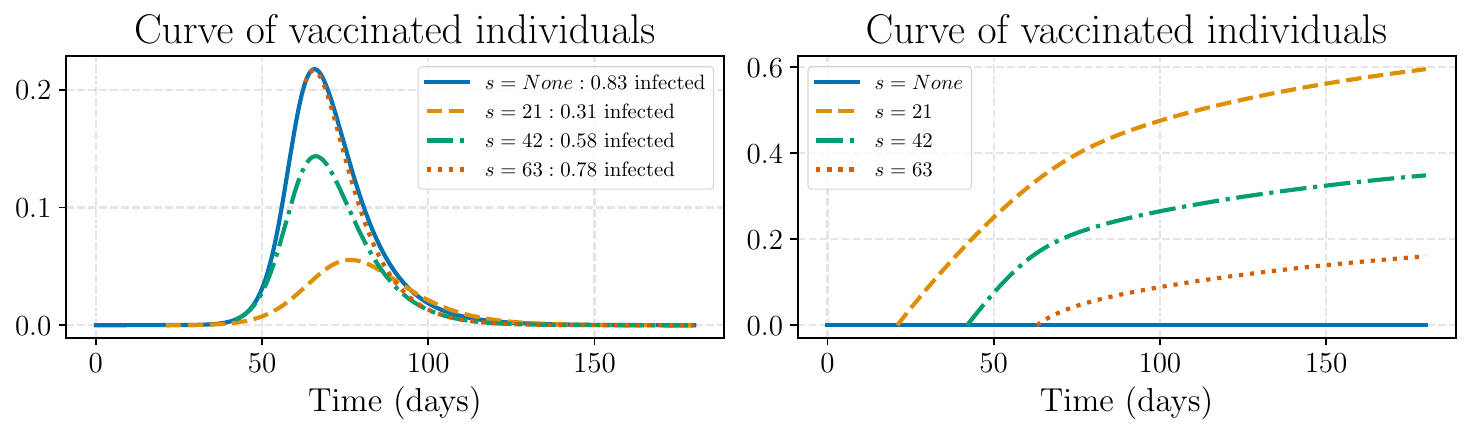}
        \caption{\label{fig:epidemic_behaviour_vaccination_different_days}Comparing different vaccination starting days}
    \end{subfigure}
    \caption{\label{fig:combined_epidemic_behaviour_vaccination}\textbf{Epidemic behavior under vaccination strategies}: 
    (a) This subfigure illustrates four epidemic scenarios under a constant vaccination strategy applied uniformly across all cities with the red line indicating the beginning of the vaccination campaign on day $s=42$. 
    The curves represent the proportion of infectious individuals (left) and the proportion of vaccinated individuals (right) across the entire metropolitan area. 
    (b) Using the same settings, this subfigure shows different starting days of the vaccine campaign for $s= 21, 42, 63$ as the first day of vaccination. 
    The parameters are: $\beta = (0.4, 0.3, 0.15, 0.15, 0.1)$, $\alpha = 0.64$, $p_{k1} = 0.2$ for each city $k > 1$, and $n = 10^{5}(50, 10, 10, 1,1)$ is the vector of population sizes, starting with only one exposed individual in the capital.}
\end{figure}

Although $\rzero$ provides a good understanding of the impact of vaccination on epidemic dynamics, it only informs about the asymptotic behavior of the dynamics. For a finite time horizon, analyzing the peak size of the epidemic provides crucial insights into the optimal vaccination strategy. 
This is the goal of \autoref{fig:epidemic_behaviour_vaccination}, which presents four different scenarios under a uniform vaccination policy. 
We have chosen $s=42$ days for this experiment. The peak size of the baseline case, when no vaccine is introduced, is about 21\% of the population. 
To reduce it to around 15\%, as shown in orange, over 60\% of the population needs to be vaccinated. 
Early vaccination can significantly reduce the peak size of the epidemic, as illustrated in \autoref{fig:epidemic_behaviour_vaccination_different_days}, which shows the impact of the vaccination starting date on the peak size.

\subsection{Evaluating the impact of the nighttime proportion \texorpdfstring{$\alpha$}{alpha}}

\begin{figure}[!htbp]
    \centering
    \includegraphics[width=\textwidth]{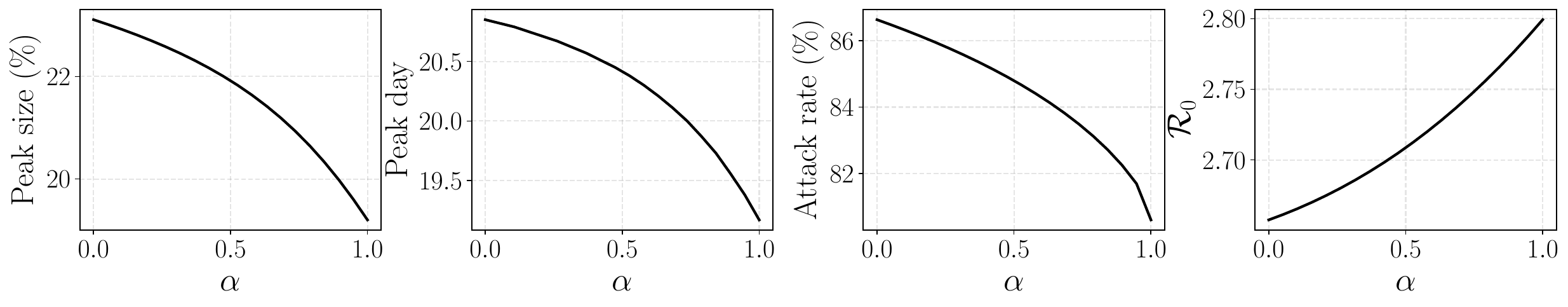}
    \caption{\label{fig:impact_alpha_epidemic}\textbf{Impact of $\alpha$ on the epidemic}: We analyze four different epidemiological quantities considering the aggregated population, the peak size (the highest value in the infectious curve), the peak day (the day when the peak size occurs), the attack rate (the number of individuals who contract the disease after $T=200$ days), and $\rzero$, as functions of $\alpha$. 
    The parameters used are based on the setting in \autoref{table:summary_scenarios_matrix_P}, except for matrix $P$, where we assume that $20\%$ of each city's population goes to the capital while the remaining population stays in their home city.}
\end{figure}

The parameter $\alpha$, representing the average proportion of the time spent in their home city, was previously shown to not have a significant impact on $\rzero$ in a two-city model.
However, in the multi-city framework, its effect on $\rzero$ is more pronounced, as depicted in \autoref{fig:impact_alpha_epidemic}.
Interestingly, an increase in $\alpha$ generates higher values for $\rzero$ but does not lead to a higher peak size or attack rate.
Another conclusion we draw is that the intensity of the epidemic increases with a higher amount of time spent in the capital, which occurs when $\alpha$ is closer to $0$.

\section{Further remarks and possible extensions}\label{sec:extensions}

\subsection{Generalizing commuting patters}

It is possible to consider different commuting times for each pair of cities $\alpha_{ij},$ which is a set of data that can be easily retrieved. One can in fact generalize this feature by taking time-dependent parameters $\alpha_{ij}$, leading to the following modified model
\[
P_i^{\mathrm{eff}}(t) = \sum_{j=1}^K (1-\alpha_{ji}(t)) p_{ji} n_j,
\]
\[
I_j^{\mathrm{eff}}(t) = \frac{1}{P_j^{\mathrm{eff}}(t)} \sum_{k=1}^K (1-\alpha_{kj}(t))  p_{kj} I_k n_k.
\]
Then, we get
\[
\frac{dS_i}{dt}(t) = \mu - \beta_i S_i \sum_{j=1}^K \alpha_{ji}(t) p_{ij}  I_i -  S_i \sum_{j=1}^K \beta_j (1-\alpha_{ij}(t)) p_{ij} I_j^{\mathrm{eff}} - \mu S_i, \]
This version also opens the door to considering infections during commuting.


\subsection{Modeling possible vaccination in the workplace}

The model presented in this paper, in particular the system~\eqref{eq:robot-dance-model-vaccination}, assumes that individuals are vaccinated in their city of residence, leading to a vaccination rate of $u_i(t) S_i(t)$ per unit of time in each city. 
However, this assumption may not hold in all cases, specifically in cities like Rio de Janeiro where the health system is universal and people can be vaccinated in any city of preference.
A particular case is when the individuals are vaccinated close to their workplace, which can be addressed as a simple modification.

To accommodate this fact, we propose an extension to our model: a proportion of individuals, denoted by $q$, get vaccinated in their city of work instead of their city of residence.
Therefore, the quantity $q p_{ij} S_i$ represents the proportion of people who live in the city $i$, work in the city $j$, and choose to be vaccinated at their workplace. 
This leads to a revised effective vaccination rate, expressed as:
\[
(1-q)u_i S_i + q\sum_{j=1}^K u_j p_{ij} S_i.
\]
This adjustment yields a modified SIR model with an altered counter variable $V_i$ represented as follows:

\begin{equation*}
    \begin{split}
        \frac{dS_i}{dt} &= \mu -\alpha \beta_i S_i I_i - (1-\alpha) S_i \sum_{j=1}^K \beta_j p_{ij} I_j^{\mathrm{eff}} - (1-q) u_i S_i - q \sum_{j=1}^K   u_j  p_{ij} S_i - \mu S_i, \\
        \frac{dI_i}{dt} &= \alpha \beta_i S_i I_i + (1-\alpha) S_i \sum_{j=1}^K \beta_j p_{ij} I_j^{\mathrm{eff}} - \gamma I_i - \mu I_i, \\
        \frac{dR_i}{dt} &= (1-q) u_i S_i + q \sum_{j=1}^K   u_j  p_{ij} S_i + \gamma I_i - \mu R_i, \\
        \frac{dV_i}{dt} &= (1-q) u_i S_i + q \sum_{j=1}^K   u_j  p_{ij} S_i.
    \end{split}
\end{equation*}

This change does not alter the paper's results regarding the overall behavior of the solution which focuses on the capital.
Considering the role of the capital as the economic hub of the metropolitan area, this model reinforces the importance of prioritizing the vaccine delivery to the capital.
This is because more people are there during the day and a higher vaccination rate effectively increases the vaccinated population in the capital and, indirectly, in the other cities.

\end{document}